\documentclass[preprint,12pt]{elsarticle}
\usepackage{amsmath,amssymb} % you already load these; keep them
\usepackage{bm}              % for \bm{...}
\usepackage{multirow}        % for \multirow in tables
\usepackage{booktabs}        % for \toprule \midrule \bottomrule
\usepackage{amsmath, amssymb}
\usepackage{graphicx}
\usepackage{lineno}
\usepackage{hyperref}
\usepackage{geometry}
\usepackage{float}
\makeatletter
\def\ps@pprintTitle{%
 \let\@oddhead\@empty
 \let\@evenhead\@empty
 \let\@oddfoot\@empty
 \let\@evenfoot\@oddfoot
}
\makeatother

\journal{Mechanical Systems and Signal Processing}

\begin{document}

\begin{frontmatter}

\title{Data-driven, Wavelet-based Identification and Reduced-order Modeling of Linear Systems with Closely Spaced Modes}

\author[mech]{Anargyros Michaloliakos}
\author[aero,navair]{Benjamin J. Chang}
\author[aero]{Lawrence A. Bergman}
\author[mech]{Alexander F. Vakakis}

\address[mech]{Department of Mechanical Science and Engineering, University of Illinois Urbana-Champaign, Urbana, IL 61801, USA}
\address[aero]{Department of Aerospace Engineering, University of Illinois Urbana-Champaign, Urbana, IL 61801, USA}
\address[navair]{Naval Air Warfare Center Aircraft Division (NAWCAD), Lexington Park, MD 20653, USA}

\begin{abstract}
This work presents a purely data-driven, wavelet-based framework for modal identification and reduced-order modeling of mechanical systems with assumed linear dynamics characterized by closely spaced modes with classical or non-classical damping distribution. Traditional Fourier-based methods often fail to reliably identify closely spaced modes or accurately capture modal interactions and complexities. To address these limitations, we propose a methodology leveraging the enhanced time–frequency resolution capabilities of the continuous wavelet transform (CWT). By selecting appropriate harmonic regions within the wavelet spectra, we effectively isolate modes, and then invert them back in the temporal domain by applying the inverse CWT (ICWT). In this way we reconstruct the corresponding modal dynamics in the time domain. Using the Hilbert transform, instantaneous phases are extracted for each identified mode, enabling the introduction of a complexified modal matrix which robustly characterizes the system's modal properties, even under challenging perturbations such as noise and uncertainties due to modal interference and unmodeled effects. The identified modal parameters are utilized to reconstruct the frequency response functions (FRFs) of the system and to develop a reduced-order model (ROM) that captures accurately the system's dominant dynamical behavior valid in a specified frequency range.. Validation of the methodology is conducted both with a numerical non-classical damping and an experimental testbed representing a model of an airplane structure. Results demonstrate the effectiveness of the proposed approach in resolving intricate modal interactions and accurately reproducing the dynamic response of complex structural systems. 
\end{abstract}


\end{frontmatter}

\newpage
\section{Introduction}

In the analysis of dynamical systems, identifying the underlying dominant time-scales governing their responses is often critical. System identification addresses precisely this challenge, focusing on the construction of accurate mathematical reduced-order models by postprocessing experimentally- or numerically-obtained system responses \cite{Reynders}. Traditional
system identification techniques typically are based on operational modal analysis in the
frequency domain such as $H_{1}, H_{2}, \text{and }H_{v}$ Frequency Response Function (FRF) estimators, or on Fourier analysis based methods such as Fast Fourier transforms (FFTs) \cite{ewins,avitable}. However, these classical frequency-domain methods inherently assume stationary behavior and thus encounter difficulties when dealing with complex, transient, or non-stationary responses.

Specifically, traditional Fourier analysis has fundamental limitations in resolving closely spaced modes and transient modal interactions \cite{Chang2025}. When two or more modes have close natural frequencies, their frequency-domain peak responses often overlap, especially under conditions of non-classical damping, incomplete available data, e.g., short time histories, or noise, e.g., for low-level ambient excitation, making modal identification difficult \cite{Qu}. This overlap  combined with the fixed frequency resolution of the Fourier transform hinder precise separation of near-frequency components \cite{hwang2003wavelet}. In large flexible structures, e.g., suspension bridges, slender buildings, and high aspect ratio wings or in structures with symmetries, such as circular rotary bladed disks (which are typical in turbomachinery), there exist high modal densities, so multiple normal modes within a narrow frequency band can produce complicated effects, e.g., combined peaks in frequency domain plots \cite{li2024review}. In these cases, classical FFT-based methods may miss or misidentify groups of closely spaced modes. Furthermore, the Fourier transform uses global trigonometric basis functions which assume signal stationarity, and, as a result, cannot capture local time-varying features in dynamic signals that occur during modal interactions, and, hence, yield highly non-stationary responses. These limitations motivate the use of multi-resolution time-frequency approaches that are capable of identifying closely spaced (in frequency) interacting modes to produce accurate and reliable ROMs both in time and frequency.

Empirical Mode Decomposition (EMD) \cite{huang1998emd} is a popular time–frequency  methodology that adaptively decomposes a measured signal into a set of oscillatory components called Intrinsic Mode Functions (IMFs). However, in practice, the extracted IMFs are seldom perfectly monochromatic since they contain more than one characteristic time (or frequency) scales. This shortcoming, known as mode mixing, typically appears when the original signal includes two frequency components that are closely spaced or when the amplitudes of adjacent modes differ greatly \cite{rilling2003emd}. To mitigate mode mixing, numerous extensions of EMD have been proposed. Previous works \cite{chen2014mech,eriten2013mech,kurt2012arch,kurt2014jsv} improved decomposition fidelity by introducing and tuning masking signals so that individual IMFs can be isolated in succession. Building on that idea, Moore et al. developed wavelet-bounded EMD (WBEMD) \cite{moore2018mssp,moore2018nonlinear}, which uses wavelet transforms to refine the separation of each IMF around its dominant frequency band, thereby producing better isolated components. Qin et al. \cite{qin2015advmat} proposed an output-only (operational) modal analysis methodology that modifies EMD to reduce mode mixing, while Sadhu \cite{sadhu2017jvc} combined multivariate and ensemble versions of EMD to obtain scale-independent IMFs. The challenge of mixed or non-separable scales is not limited to EMD though it is pervasive in engineering and the applied sciences whenever responses become highly non-stationary or dynamically complex , e.g., near bifurcations, during abrupt dynamical transitions or when systems lose synchronicity and new frequency components emerge. The works of Ma et al. \cite{ma2009decoupling, ma2010decoupling, salsa2021advances} decouple the damped equations of motion of discrete and continuous vibrating systems and provide mathematical formulations that, in principle, could be used in system identification; however, these results are not used in this work as explained below.

The primary objective of this study is to introduce a broadly applicable data-driven wavelet-based framework for modal identification of linear time-invariant mechanical systems, capable of extracting modal parameters, i.e., natural frequencies, damping ratios, and complex-valued mode shapes, directly from output-only measurements. The approach is built around wavelet analysis and is especially useful in situations where traditional modal analysis techniques become ineffective, such as when dealing with closely spaced modes, strong modal coupling, and/or systems with non-classical damping. This framework begins with the continuous wavelet transform (CWT) to identify key harmonic components, and the inverse CWT (ICWT) to generate isolated harmonic component time histories. Subsequently, amplitude envelopes are estimated using the Hilbert transform and combined with a new strategy for phase estimation to accurately compute complex-valued mode shapes. This allows for reliable identification of modal properties even in the presence of modal interactions or transient effects. One of the strengths of this framework is that it is entirely data-driven; that is, it does not require assumptions about system models or input measurements. Moreover, the mode shapes obtained are automatically mass-orthonormalized, which enables immediate comparison to theoretical results or finite element simulations. The time–frequency resolution offered by the wavelet-based framework also makes it particularly effective for studying interesting dynamics such as beating phenomena, highly non-stationary, transient responses, and other complex dynamic behaviors that might be missed by traditional tools.

The paper is structured as follows. Section~2 provides the theoretical background necessary for the proposed method, detailing fundamental concepts related to the CWT and ICWT, Hilbert transform, FRF reconstruction, as well as a comprehensive, step-by-step outline of the proposed system identification protocol. In Section~3, the efficacy of the approach is illustrated through its application to a numerical multiple degree of freedom (MDOF) system exhibiting closely spaced modes and non-classical damping, accompanied by thorough validation via FRF reconstruction and reduced-order modeling. Section~4 further demonstrates the practicality of the developed methodology through experimental validation performed on a laboratory-scale airplane model, where modal parameters are accurately extracted from measured responses. Finally, Section~5 summarizes the main contributions and achievements of the study, and highlights promising avenues for future research in the field of wavelet-based structural system identification.

\section{Postprocessing Tools, Analytical Concepts, and Data-Driven Methodology}

In this section, we discuss key postprocessing tools and analytical concepts, including CWT and ICWT, Hilbert transforms, and FRF reconstructions, which form the basis of the proposed identification protocol. This theoretical framework supports the robust extraction of modal parameters directly from measured data, enabling subsequent application and validation using both numerical and experimental measurements. We note that, even if the postprocessing tools and analytical methods reviewed below are not new, we still briefly review them in order to provide the necessary rationale for their integration into the proposed data-driven methodology.

\subsection{Continuous Wavelet Transform (CWT)}

The CWT is a powerful tool for mapping signals from the time domain into a joint time–frequency domain representation. Unlike the traditional Fourier transform, which provides a broad overview of \textit{stationary} frequency content, the CWT is effective in resolving local and time-varying features in an adaptive way, making it ideal for studying even highly non-stationary signals with multiple varying frequencies and complex (nonlinear) dynamics. For a given signal \( x(t) \), its corresponding CWT is expressed as the following complex quantity,

\begin{equation}
X(a, b) = \frac{1}{\sqrt{a}}\int_{-\infty}^{\infty} x(s)\,\overline{\psi}\left(\frac{s - b}{a}\right) ds,
\label{eq:cwt_ab}
\end{equation}

\noindent where \( \psi(t) \) is a complex-valued wavelet function localized in both time and frequency, \( a \) denotes a scaling parameter (inversely classical to frequency), and \( b \) a time shift parameter. By relating the scale parameter \( a \) directly to angular frequency \(\omega\), through the relation \( a = \frac{\omega_c}{\omega} \) (where \( \omega_c \) is the center frequency of the wavelet), Eq.~\eqref{eq:cwt_ab} can be reformulated explicitly in the frequency domain as,

\begin{equation}
X(\omega, t) = \sqrt{\frac{\omega}{\omega_c}}\int_{-\infty}^{\infty} x(s)\,\overline{\psi}\left(\frac{\omega(s - t)}{\omega_c}\right) ds.
\label{eq:cwt_omega}
\end{equation}
The center frequency \( \omega_c \) of a wavelet transform is determined by its frequency content through the following expression,

\begin{equation}
\omega_c = \frac{\int_{0}^{\infty}\omega^2|\Psi(\omega)|^2 d\omega}{\int_{0}^{\infty}|\Psi(\omega)|^2 d\omega},
\label{eq:centerfreq}
\end{equation}

\noindent where \( \Psi(\omega) \) denotes the Fourier transform of the wavelet \( \psi(t) \). To qualify as a wavelet function suitable for the CWT, two critical mathematical conditions must be satisfied. First, the wavelet must be an integrable function, i.e., possess finite energy,

\begin{equation}
\int_{-\infty}^{\infty}|\psi(t)|^2 dt < \infty,
\label{eq:finite_energy}
\end{equation}

\noindent and second, it must fulfill the admissibility condition,

\begin{equation}
C = \int_{0}^{\infty}\frac{|\Psi(\omega)|^2}{\omega} d\omega < \infty.
\label{eq:admissibility}
\end{equation}
Additionally, for certain wavelets such as the Morlet wavelet, a supplementary condition ensures zero mean at zero frequency,

\begin{equation}
\lim_{\omega \to 0}\mathrm{Re}\{\Psi(\omega)\} = 0, \quad \lim_{\omega \to 0}\mathrm{Im}\{\Psi(\omega)\} = 0.
\label{eq:morlet_zero}
\end{equation}

The Morlet wavelet, widely utilized in structural dynamics and vibration analysis due to its excellent capacity to separate phase and amplitude information, is defined explicitly as,

\begin{equation}
\psi(t) = \pi^{-1/4}\left(e^{i\omega_c t} - e^{-\omega_c^2/2}\right)e^{-t^2/2}.
\label{eq:morlet_wavelet}
\end{equation}
Substituting this wavelet into Eq.~\eqref{eq:cwt_omega}, the explicit analytic expression of the Morlet CWT:

\begin{equation}
X(\omega, t) = \sqrt{\frac{\omega}{\sqrt{\pi}\omega_c}}\int_{-\infty}^{\infty} x(s)\left(e^{-i\omega(s - t)} - e^{-\omega_c^2/2}\right)e^{-\frac{1}{2}\left(\frac{\omega(s - t)}{\omega_c}\right)^2}ds.
\label{eq:morlet_cwt}
\end{equation}
Although Eq.~\eqref{eq:morlet_cwt} offers an exact representation of the CWT, its direct numerical determination is computationally intensive for large data sets. A more efficient approach arises by interpreting the CWT as a convolution between the original signal \( x(t) \) and a scaled and shifted wavelet \( \psi(t) \). Leveraging Parseval's theorem, this convolution-based representation can be efficiently implemented via the Fourier transform as follows,

\begin{equation}
X(\omega, t) = \sqrt{\frac{\omega_c}{\omega}}\int_{-\infty}^{\infty}\hat{x}(\xi)\,\overline{\Psi}\left(\frac{\xi\omega}{\omega_c}\right)e^{i\xi t} d\xi,
\label{eq:cwt_fft}
\end{equation}

\noindent where \( \hat{x}(\xi) \) represents the Fourier transform of the signal \( x(t) \). This frequency-domain interpretation is particularly advantageous, as it enables fast and computationally effective implementations of the CWT through efficient numerical algorithms, such as the FFT. This computational framework is the cornerstone of the data-driven system identification methodology of this study.

\subsection{Inverse Continuous Wavelet Transform}

Following an inverse path, to recover the original signal \( x(t) \) from its wavelet representation \( X(\omega, t) \), one can apply the ICWT \cite{MOJAHED2021}. Using the frequency-based formulation in Eq.~\eqref{eq:cwt_omega}, the ICWT is expressed as,

\begin{equation}
x(t) = -\frac{1}{\omega_c C} \int_0^\infty \int_{-\infty}^\infty X(\omega, s) 
\sqrt{\frac{\omega}{\omega_c}} \psi \left( \frac{\omega(t - s)}{\omega_c} \right) ds\, d\omega,
\label{eq:icwt_direct}
\end{equation}

\noindent where \( C \) is the wavelet admissibility constant and \( \omega_c \) is the wavelet center frequency. While Eq.~\eqref{eq:icwt_direct} provides an exact reconstruction, it is computationally expensive for practical applications. To address this, we again use Parseval’s theorem to derive an efficient frequency-domain formulation of the ICWT,

\begin{equation}
x(t) = \frac{1}{\omega_c C} \int_0^\infty \int_{-\infty}^\infty 
\tilde{X}(\omega, \xi) \bar{\Psi} \left( \frac{\xi \omega}{\omega_c} \right) e^{i \xi t} d\xi\, d\omega,
\label{eq:icwt_fft}
\end{equation}

\noindent where \( \tilde{X}(\omega, \xi) \) is the Fourier transform of \( X(\omega, t) \) with respect to its time variable \( t \), and \( \bar{\Psi} \left( \frac{\xi \omega}{\omega_c} \right) \) is the Fourier transform of the wavelet. The representation in Eq.~\eqref{eq:icwt_fft} is more amenable to FFT and inverse FFT (IFFT) based implementations and significantly accelerates computational efficiency.

\medskip
To reconstruct the signal using only specific harmonic components, the CWT is partitioned into \( N \) "harmonic regions" in the time-frequency plane \cite{MOJAHED2021}. The signal \( X(\omega, t) \) is thus decomposed as,

\begin{equation}
X(\omega, t) = \sum_{j=1}^N X_j(\omega, t),
\label{eq:decomp_sum}
\end{equation}

\noindent where \( X_j(\omega, t) \) denotes the portion of the wavelet spectrum associated with the \( j \)-th harmonic region. Each region is defined using frequency thresholds \( \omega_j(t) \) as follows,

\begin{equation}
X_j(\omega, t) = X(\omega, t) \left[ H(\omega - \omega_{j-1}(t)) - H(\omega - \omega_j(t)) \right],
\label{eq:region_decomp}
\end{equation}

\noindent where \( H(\cdot) \) is the Heaviside step function. This formulation effectively creates non-uniform frequency strips in the CWT spectrum, isolating specific harmonic components.

The recovered time-domain signal is expressed as the sum of \( N \) harmonic contributions,

\begin{equation}
x(t) = \sum_{j=1}^N x_j(,t),
\label{eq:sum_harmonics}
\end{equation}

\noindent where each harmonic component \( x_j(t) \) is reconstructed independently from its corresponding region via:

\begin{equation}
x_j(t) = -\frac{1}{\omega_c C} \int_0^\infty \int_{-\infty}^\infty 
X_j(\omega, s) \sqrt{\frac{\omega}{\omega_c}} 
\psi \left( \frac{\omega(t - s)}{\omega_c} \right) ds\, d\omega.
\label{eq:icwt_xj}
\end{equation}

\noindent Alternatively, this can again be computed efficiently in the frequency domain as,

\begin{equation}
x_j(t) = \frac{1}{\omega_c C} \int_0^\infty \int_{-\infty}^\infty 
\tilde{X}_j(\omega, \xi) \bar{\Psi} \left( \frac{\xi \omega}{\omega_c} \right) e^{i \xi t} d\xi\, d\omega,
\label{eq:icwt_xj_fft}
\end{equation}

\noindent where \( \tilde{X}_j(\omega, \xi) \) is the Fourier transform of \( X_j(\omega, t) \) with respect to time. This harmonic decomposition approach offers significant flexibility -- users can define tailored regions to precisely isolate desired modal responses depending on the specific features of the non-stationary measured time series (e.g., harmonic components with highly varying frequencies in time due to nonlinear effects). Such decomposition is especially valuable for analyzing systems with mode mixing, time-varying characteristics, or non-stationary features, as will be demonstrated in later sections.
\subsection{Hilbert Transform}

The Hilbert transform is a fundamental tool in signal analysis, particularly useful for extracting instantaneous amplitude, phase, and frequency from real-valued time-domain, monochromatic (or near-monochromatic) signals. It is especially relevant in the context of modal analysis, where closely spaced modes and non-classical damping are often realized.

Given a real-valued signal \( x(t) \), its Hilbert transform \( \mathcal{H}\{x(t)\} \) is defined as,

\begin{equation}
\mathcal{H}\{x(t)\} = \frac{1}{\pi} \, \text{P.V.} \int_{-\infty}^{\infty} \frac{x(\tau)}{t - \tau} d\tau,
\label{eq:hilbert_def}
\end{equation}

\noindent where "P.V." denotes the Cauchy principal value of the integral. The result is a 90-degree phase-shifted version of \( x(t) \), forming the imaginary part of the analytic signal:

\begin{equation}
z(t) = x(t) + i \, \mathcal{H}\{x(t)\}.
\label{eq:analytic_signal}
\end{equation}

The analytic signal \( z(t) \) can be used to extract important time-varying characteristics of the signal. Indeed, the instantaneous amplitude \( A(t) \) and instantaneous phase \( \phi(t) \) of the signal are defined as,

\begin{equation}
A(t) = |z(t)| = \sqrt{x^2(t) + \mathcal{H}\{x(t)\}^2}, \qquad
\phi(t) = \arg(z(t)) = \tan^{-1} \left( \frac{\mathcal{H}\{x(t)\}}{x(t)} \right).
\label{eq:amp_phase}
\end{equation}

This transform is particularly useful for analyzing non-stationary or modulated signals, such as those arising in structures with closely spaced or time-varying modes. In the present study, the Hilbert transform is employed as a post-processing tool to refine the interpretation of wavelet-derived modal signals, allowing for complexification of the modal matrix estimate as shown in the following sections.

\subsection{Frequency Response Functions}

An FRF characterizes the steady state dynamics of a linear time-invariant (LTI) system under external excitation in the frequency domain, as it provides a direct relationship (transfer function) between the input and output for this system. As such, it is a foundational tool in system identification, vibration analysis, and modal testing.

For a single-input, single-output (SISO) system generally described by the second-order differential equation,

\begin{equation}
M \ddot{x}(t) + C \dot{x}(t) + K x(t) = f(t),
\end{equation}

\noindent where \( M \), \( C \), and \( K \) are the mass, damping, and stiffness matrices, respectively, the system's frequency response to a harmonic input \( f(t) = F_0 e^{i\omega t} \) where $\omega$ is the cyclic frequency (in rad/s), and $\omega = 2\pi f$, where $f$ is the cyclic frequency (in Hz) can be written as,

\begin{equation}
X(\omega)e^{i\omega t} = H(\omega) F_0 e^{i\omega t},
\end{equation}

\noindent where \( H(\omega) \) is the FRF matrix, defined as,

\begin{equation}
H(\omega) = \left[ -\omega^2 M + i\omega C + K \right]^{-1}.
\label{eq:frf}
\end{equation}

\noindent In the expressions above, $\omega$ is cyclic frequency (in rad/sec), and is expressed as, $\omega = 2\pi f$ where $f$ is the cyclic frequency (in Hz). Each entry \( H_{ij}(\omega) \) describes the frequency response at degree of freedom \( i \) due to excitation at degree of freedom \( j \). Drive point FRFs, i.e., FRFs corresponding to degrees of freedom (DOFs) that are forced, lie on the diagonal ($i=j$) while indirect (not directly driven) points, i.e., FRFs corresponding to DOFs that are not forced, lie on off diagonal terms ($i\neq j$). The FRF matrix is symmetric for LTI systems whose dynamics is referred with respect to an inertial coordinate frame.

In practice, the input-output data used to compute FRFs are typically generated by applying known excitation at a point using an impact hammer or shaker and measuring the response at specific points using accelerometers. The ratio of the Fourier transform of the response signal to that of the input signal yields an experimental estimate of the FRF,

\begin{equation}
H(\omega) = \frac{X(\omega)}{F(\omega)},
\end{equation}

\noindent where \( X(\omega) \) and \( F(\omega) \) are the Fourier transforms of the response and forcing signals, respectively. Other common data-driven FRF forms include the $H_{1}, H_{2}, \text{and } H_{v}$ estimators \cite{ewins,avitable}.The $H_1$ estimator minimizes the effect of output noise and is given by
\[
H_1(\omega) = \frac{G_{XF}(\omega)}{G_{FF}(\omega)},
\]
while the $H_2$ estimator minimizes the effect of input noise and is given by
\[
H_2(\omega) = \frac{G_{XX}(\omega)}{G_{FX}(\omega)}.
\]
Here, \( G_{AB}(\omega) \) denotes the cross- or auto-spectral density between signals \( A \) and \( B \). The \( H_v \) estimator is a vector-averaged version that combines features of both \( H_1 \) and \( H_2 \), offering improved performance under certain noise conditions.

FRFs are instrumental in modal analysis, as their peaks, phase crossings, and overall shapes are used to extract natural frequencies, damping ratios, and mode shapes via modal analysis \cite{ewins}, and curve-fitting algorithms such as PolyMAX \cite{Peeters}. In the present study, FRFs are used as a reference to validate the accuracy of the wavelet-based modal decomposition and the resulting reduced-order model.

\subsection{Wavelet-Based System Identification}
\label{sec:wavelet_protocol}

\begin{figure}[htbp]
    \centering
    \includegraphics[width=0.45\textwidth]{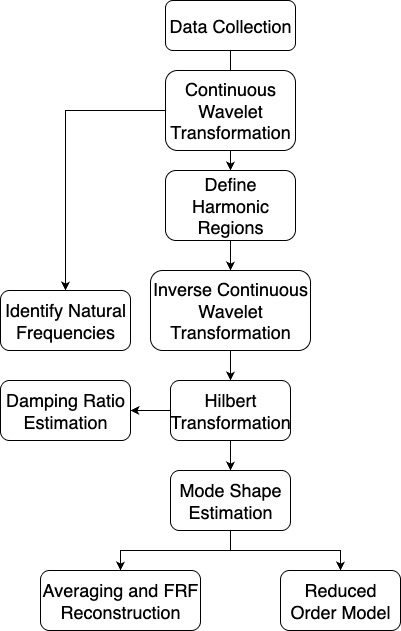}
    \caption{Wavelet-based system identification protocol.}
    \label{fig:flowchart}
\end{figure}

Integrating and building upon the postprocessing tools and theoretical foundations reviewed in the previous sections, we now detail an integrated protocol for the data-driven, wavelet-based system identification applied to dynamic systems with closely spaced modes with classical or non-classical damping distributions. The primary goal of this method is to extract accurate estimates for the system modal parameters, i.e., natural frequencies, damping ratios, and complex-valued mode shapes, directly from measured time-domain responses, without dependence on predetermined model parametric assumptions. The full identification workflow is summarized schematically in the flowchart provided in Fig.~\ref{fig:flowchart}.

Initially, dynamic response data are collected from the linear, time-invariant system under investigation and processed into velocity signals (if not directly measured). The CWT, as defined by Eq.~\eqref{eq:cwt_omega}, is then applied to these measured velocity time histories, resulting in a detailed joint time–frequency representation. This representation makes it much easier to identify closely spaced modes, capture non-stationary transient events, and understand how energy is distributed across frequencies - aspects that are often difficult to detect with traditional Fourier-based methods. Accurately identifying natural frequencies, however, is highly dependent on wavelet parameter selection. In particular, using a higher mother wavelet parameter can significantly improve frequency resolution, allowing for more precise detection of modal frequencies. As demonstrated in prior studies (e.g., \cite{MOJAHED2021}), this approach also permits the tracking of harmonic component evolution over time, proving particularly useful when addressing nonlinear phenomena such as softening or hardening stiffness effects. Hence, the versatility of the wavelet transformations is exploited by selecting wavelet parameters optimized for high frequency resolution, ensuring reliable modal frequency estimation.

After construction of the CWT spectra, and without relying on any prior knowledge of the system itself, we apply the following inverse wavelet harmonic decomposition methodology for modal identification. For each measured velocity signal \(\dot{x}_i(t)\), \(i = 1, \dots, N\), we begin by identifying the harmonic regions of interest within the CWT representation in the time–frequency domain. These harmonic regions are selected based on the observed topology of the CWT spectrum. In particular, the boundaries of these regions are chosen to separate the components of the response corresponding to different (suspected) modes. As discussed in \cite{MOJAHED2021}, a typical approach is to first identify the harmonic regions from the single response \(\dot{x}_k(t)\) with the richest spectral content, and then apply the same harmonic boundaries to all remaining DOFs \(i \ne k\). This approach is justified because the harmonic signatures of each mode are expected to persist across all DOFs due to the distributed nature of modal dynamics.

We then numerically apply the ICWT expression \eqref{eq:icwt_direct} to extract the \(j\)-th harmonic component of each velocity response \(\dot{x}_i^{(j)}(t)\), for \(j = 1, \dots, N\), corresponding to the \(N\) considered harmonic regions. This process yields a time-domain representation of each individual mode's contribution to each response signal.
To obtain the moduli of the $j$-th orthonormalized mode shape, we consider the amplitude of the decomposed harmonic component $\dot{x}_k^{(j)}$, denoted by its time-averaged envelope $\left\langle \dot{x}_k^{(j)} \right\rangle$. The normalized modal modulus corresponding to the $k$-th DOF is then given by,
\begin{equation}
    \left\langle \varphi_j \right\rangle_k = \frac{\left\langle \dot{x}_k^{(j)} \right\rangle}{M_j},
\end{equation}
where the normalization factor $M_j$ is computed as \cite{MOJAHED2021}:
\begin{equation}
    M_j = \left[ \sum_{k=1}^N \left\langle \dot{x}_k^{(j)} \right\rangle^2 \right]^{1/2}.
\end{equation}
This step yields mass-orthonormalized moduli for our estimations of the modal matrix and forms the foundation for subsequent mode shape complexification and reconstruction procedures.

Significant focus is then placed on accurately estimating the phase difference between matching harmonic components of each DOF. This is necessary to account for the possibility of modal complexity, which is anticipated in closely spaced modes with non-classical damping. Following extraction of modal amplitudes, complex-valued mode shapes are introduced by applying the Hilbert transform, Eq.~\eqref{eq:amp_phase}, to obtain the corresponding instantaneous phases. To this end, a reference DOF is selected and the instantaneous phase difference \(\theta_k\) for each remaining DOF is computed with respect to the reference DOF, and is combined with the corresponding moduli to construct the complex-valued mode shape vector \(\boldsymbol{\Psi}_j\):

\begin{equation}
\boldsymbol{\Psi}_j = 
\begin{bmatrix}
 \left\langle \varphi_1 \right\rangle e^{i \theta_1} \\
 \left\langle \varphi_2 \right\rangle e^{i \theta_2} \\
 \vdots \\
 \left\langle \varphi_N \right\rangle e^{i \theta_N}
\end{bmatrix},\quad
\boldsymbol{\Phi}_j =
\begin{bmatrix}
 \left\langle \varphi_1 \right\rangle \\
 \left\langle \varphi_2 \right\rangle \\
 \vdots \\
 \left\langle \varphi_N \right\rangle
\end{bmatrix},
\end{equation}

\noindent where \(\left\langle \varphi_k \right\rangle\) denotes the amplitude (modulus) of the modal participation of the \(k\)-th DOF.

Lastly, to estimate the modal damping ratios, we utilize the decomposed harmonic components. Specifically, the modal damping ratio $\zeta_k$ is derived by fitting an exponential decay to the instantaneous amplitude envelope and extracting the slope of the logarithmic amplitude decay. This approach allows for robust damping estimation directly from measured response data, but other approaches such as the concept of equivalent damping described by Remick et al.  can be used as well \cite{REMICK2014}.

To obtain accurate phase estimates, we use a time windowing approach. The objective is to select overlapping windows that meet two key conditions: (i) the signal’s envelope within each window should be smooth and have sufficient magnitude (strength of signal) and (ii) the window must include a time segment shared by both the reference DOF and the one being analyzed. This reduces the effects of noise and prevents inconsistencies in phase identification. Different regions of the structure might be excited by the applied load(s) non-uniformly, so phase estimation is performed across several measurement sets. Finally, an optimization step combines these estimates, averaging them in a way that is robust to variation and produces reliable complex mode shapes.

After estimating natural frequencies, damping ratios, and complex-valued mode shapes, the system FRFs are reconstructed using the extracted modal parameters through Eq.~\eqref{eq:frf_recon}. This reconstruction serves as a rigorous validation tool by enabling direct comparison with measured or simulated frequency-domain responses. Finally, a ROM is constructed using the identified modal parameters, which reproduces the dominant dynamics in the frequency range of interest, and is suitable for subsequent analysis, design optimization, and control applications.

\subsubsection{Varying the Location of the Excitation: Velocity Data Recording}
\label{sec:drive_point_cases_velocity}

Due to the potential variability in modal excitation caused by the different excitation (drive) points, obtaining accurate mode shape moduli and phase estimates requires analysis across multiple measurement data sets. In practice, due to the variability in modal participation due to the locations of excitation and measurement, each excitation point uniquely excites the modes of a structure, potentially resulting in some modes being more prominently observed at particular measurement locations, while being less visible or obscured in others. Consequently, to ensure robust and reliable modal identification, the moduli and phase estimation procedures are conducted separately for each drive point measurement.

Specifically, having recorded multiple estimates of the modal matrix $\boldsymbol{\Psi}^{(k)}$ for different drive points indexed by $k$, we introduce an optimization procedure to combine these estimates into a unified modal representation. For the $j$-th mode, we seek to determine a set of non-negative weights $\{w_j^{(k)}\}_{k=1}^{K}$ satisfying the normalization constraint $\sum_{k=1}^{K} w_j^{(k)} = 1$, such that the optimally weighted combination of modal matrices is given by,
\begin{equation}
    \boldsymbol{\psi}_j^{*} = \sum_{k=1}^{K} w_j^{(k)} \boldsymbol{\psi}_j^{(k)}.
\end{equation}

The weights are optimized by reconstructing FRFs from the combined modal matrix $\boldsymbol{\Psi}^{*}$ and minimizing the reconstruction error E,
\begin{equation}
    E = \sum_{f \in \mathcal{F}} \sum_{i=1}^{N} \left| FRF_{\text{meas}}(f,i) - FRF_{\text{recon}}(f,i;\boldsymbol{\Psi}^{*}) \right|^2,
\end{equation}
where $FRF_{\text{meas}}(f,i)$ and $FRF_{\text{recon}}(f,i;\boldsymbol{\Psi}^{*})$ represent the measured and reconstructed FRFs, respectively, over the discrete frequency set $\mathcal{F}$ at the $i$-th measurement location, and f is the cyclic frequency of the excitation. This optimization ensures that the final mode shape estimates are not biased toward any particular drive point measurement and remain representative of the overall modal properties of the system. Thus, the final complex modal matrix robustly captures the true system dynamics, accounting for variations that arise from multiple excitation scenarios.

\subsection{Validation}
To validate the accuracy and physical relevance of the modal parameters extracted via the data-driven, wavelet-based identification protocol, two complementary validation approaches are employed in this work. First, a frequency-domain comparison is performed by reconstructing the system’s FRFs using the identified modal properties and comparing the results with the measured FRFs. Second, a wavelet-based (time-frequency domain) validation is carried out by constructing a ROM based on the extracted natural frequencies, damping ratios, and mode shapes. The ROM response is simulated and compared in the time-frequency domain to the measured system output under subject to the actual input conditions. Together, these two approaches provide a rigorous assessment of the efficacy of the proposed method.

\subsubsection{Frequency Response Function Reconstruction for Validation}
\label{sec:frf_reconstruction}

To assess the validity and accuracy of the modal identification approach, we reconstruct system FRFs from the extracted modal parameters. Assuming a linear system with \( N \) modes, the \( k \)-th mode is characterized by its natural frequency \( \omega_k \), and viscous damping ratio \( \zeta_k \) summarized by the complex pole,

\begin{equation}
\lambda_k = \omega_k \left( \zeta_k - j \sqrt{1 - \zeta_k^2} \right).
\label{eq:complex_pole}
\end{equation}

In the Laplace domain, if \( F(s) \) is the Laplace transform of the force input and \( Y(s) \) is that of the displacement output, the receptance FRF (i.e., displacement output per unit force input) is given by,

\begin{equation}
H(s) = \frac{Y(s)}{F(s)}.
\end{equation}

Because \( Y(s) = \frac{V(s)}{s} \), where \( V(s) \) is the Laplace transform of the velocity, we also have,

\begin{equation}
H(s) = \frac{1}{s} \frac{V(s)}{F(s)} = \frac{1}{s} \cdot \text{Mobility}(s),
\end{equation}

\noindent and similarly, the accelerance FRFs is defined as,

\begin{equation}
\quad \text{Accelerance}(s) = \frac{A(s)}{F(s)} = s \cdot \text{Mobility}(s).
\end{equation}

To express the FRF in modal coordinates, we use the following decomposition based on the system's eigenstructure. The receptance FRF at frequency \( \Omega \) can be expressed as,

\begin{equation}
H(\Omega) = \sum_{k=1}^{N} \frac{Q_k \boldsymbol{\Psi}_k \boldsymbol{\Psi}_k^\top}{j \Omega - \lambda_k} 
+ \frac{Q_k^* \boldsymbol{\Psi}_k^* {\boldsymbol{\Psi}_k^*}^\top}{j \Omega - \lambda_k^*},
\label{eq:frf_recon}
\end{equation}

\noindent where \( \boldsymbol{\Psi}_k \) is the complex mode shape of the \( k \)-th mode, (.)* denotes complex conjugate, \( Q_k \) is a modal scaling constant, and \( \lambda_k \) is the complex eigenvalue. The second term accounts for the complex conjugate mode. This formulation yields the analytical FRF, which can be directly compared against experimentally measured or numerically-simulated FRFs for validation purposes.

\subsubsection{ROM–Wavelet Comparison}
\label{sec:rom_reconstruction}

To further assess the physical relevance of the identified modal parameters, we construct a ROM and compare simulated time histories to measured time histories. The ROM formulation is based on modal superposition, whereby the system response is projected onto the subspace spanned by the identified dominant complex mode shapes \( \boldsymbol{\Psi} \).

Selecting a specific frequency range where the system possesses \( r \) modes, the modal transformation reads,

\begin{equation}
\boldsymbol{\Psi}^T \mathbf{M} \boldsymbol{\Psi} = \mathbf{I}_r, \quad
\boldsymbol{\Psi}^T \mathbf{K} \boldsymbol{\Psi} = \boldsymbol{\Omega}^2, \quad
\boldsymbol{\Psi}^T \mathbf{C} \boldsymbol{\Psi} = 2 \boldsymbol{\Xi} \boldsymbol{\Omega},
\label{eq:modal_transform}
\end{equation}

\noindent where \( \boldsymbol{\Omega} = \text{diag}(\omega_1, \dots, \omega_r) \) contains the identified natural frequencies, and \( \boldsymbol{\Xi} = \text{diag}(\zeta_1, \dots, \zeta_r) \) contains the associated damping ratios. Note that for the general case of non-classical damping distribution, the last expression in Eq.~\eqref{eq:modal_transform} is a non-diagonal matrix (whereas the first two matrices are diagonal). Hence, we proceed to a state-space formulation.

In state space, the modal equations of motion in terms of the transformed coordinates \( \mathbf{q}(t) \) are given by,

\begin{equation}
\begin{pmatrix}
\dot{\mathbf{q}}(t) \\
\ddot{\mathbf{q}}(t)
\end{pmatrix}
=
\begin{pmatrix}
\mathbf{0} & \mathbf{I}_r \\
-\boldsymbol{\Omega}^2 & -2 \boldsymbol{\Xi} \boldsymbol{\Omega}
\end{pmatrix}
\begin{pmatrix}
\mathbf{q}(t) \\
\dot{\mathbf{q}}(t)
\end{pmatrix}
+
\begin{pmatrix}
\mathbf{0} \\
\boldsymbol{\Psi}^T
\end{pmatrix}
\mathbf{f}(t),
\label{eq:rom_state}
\end{equation}

\noindent which can be recast in standard state-space form as,

\begin{equation}
\dot{\mathbf{X}}(t) = \mathbf{A}_{\text{modal}} \, \mathbf{X}(t) + \mathbf{B}_{\text{modal}} \, \mathbf{u}(t), \qquad
\mathbf{y}(t) = \mathbf{C}_{out} \, \mathbf{X}(t),
\label{eq:rom_ss}
\end{equation}

\noindent where $\mathbf{X}(t)$ is the state vector composed of modal coordinates and velocities, $\mathbf{A}_{\text{modal}}$ and $\mathbf{B}_{\text{modal}}$ are matrices constructed from the identified modal parameters, and $\mathbf{C}_{out}$ is the output matrix used in traditional state-space forms. This represents the identified ROM for the frequency range of interest.

To validate the accuracy of the outlined data driven, wavelet-based system identification methodology, two systems were studied: (i) a three DOF spring-mass-damper system with closely spaced modes and non-classical damping, and (ii) an airplane model with closely spaced modes. For the discrete three DOF system, the methodology was applied to direct simulation data for system identification, and the results were validated by comparison of the exact and reconstructed FRFs and time-frequency responses. For the experimental airplane model, the same comparisons were made, but with measured data in lieu of exact simulation data. The following sections provide details of both studies.

\section{Numerical Study: Closely Spaced Modes with Non-classical Damping}

\subsection{System Setup and Parameters}

\begin{figure}[htbp]
    \centering
    \includegraphics[width=0.6\textwidth]{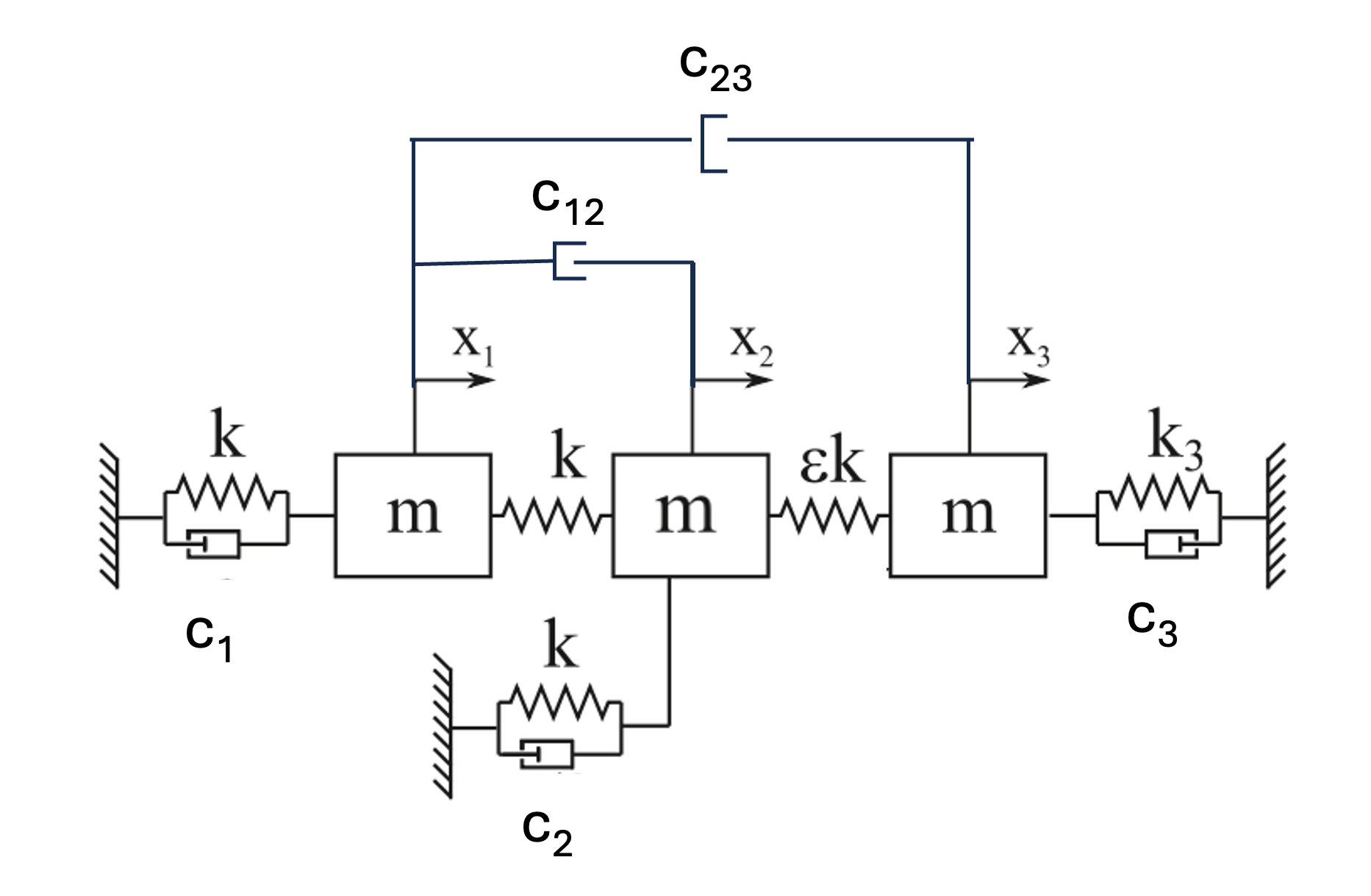}
    \caption{Schematic representation of the three coupled oscillator system with non-classical damping distribution, and closely spaced modes.}
    \label{fig:system_schematic}
\end{figure}

In the first validation study, we consider a linear three-DOF system consisting of linear, asymmetrically damped coupled oscillators, configured with weak inter-element coupling, which yields two closely spaced modes; it is known that such weak coupling and unevenly distributed damping can be a source of mode mixing and identification difficulty in data-driven approaches. The schematic representation of this system is shown in Fig.~\ref{fig:system_schematic}.

The governing equations of motion are given by,
\begin{equation}
\ddot{\bm{x}} + \mathbf{D} \dot{\bm{x}} + \mathbf{S} \bm{x} = 
\mathbf{M}^{-1}\bm{f(t)}, 
\quad \bm{x}(0) = \bm{0}, \quad \dot{\bm{x}}(0) = \bm{0}
\label{eq:gov_eq}
\end{equation}

\noindent where the mass, damping, and stiffness matrices are defined as follows,
\begin{equation}
\mathbf{M} = \begin{bmatrix}
\frac{m}{2} & 0 & 0 \\
0 & \frac{m}{2} & 0 \\
0 & 0 & \frac{m}{5}
\end{bmatrix}, \quad
\mathbf{C} = \begin{bmatrix}
c_{2A} + c_{3A} + c_1 & -c_{2A} & -c_{3A} \\
-c_{2A} & c_{2A} + \frac{c_2}{2} & 0 \\
-c_{3A} & 0 & c_{3A} + c_3
\end{bmatrix}, 
\end{equation}

\begin{equation}
\mathbf{K} = \begin{bmatrix}
2k & -k & 0 \\
-k & (2 + \epsilon)k & -\epsilon k \\
0 & -\epsilon k & k_3 + \epsilon k
\end{bmatrix}.
\end{equation}

\begin{figure}[H]
    \centering
    \includegraphics[width=1\textwidth]{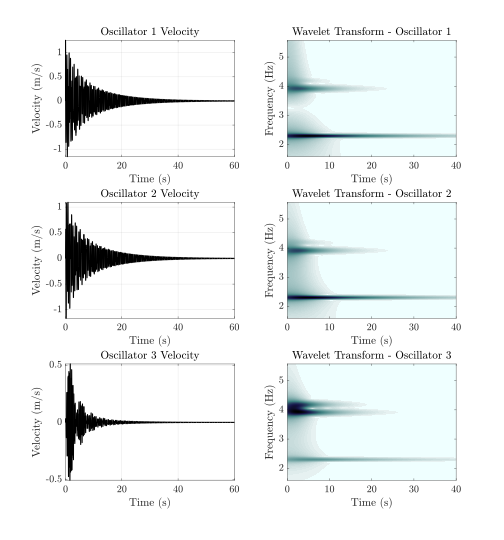} 
    \caption{Drive point at oscillator 1; time-frequency representation of the system response using the CWT. The beat phenomenon between the two closely spaced modes is clearly visible.}
    \label{fig:cwt_response_1}
\end{figure}

\begin{figure}[H]
    \centering
    \includegraphics[width=1\textwidth]{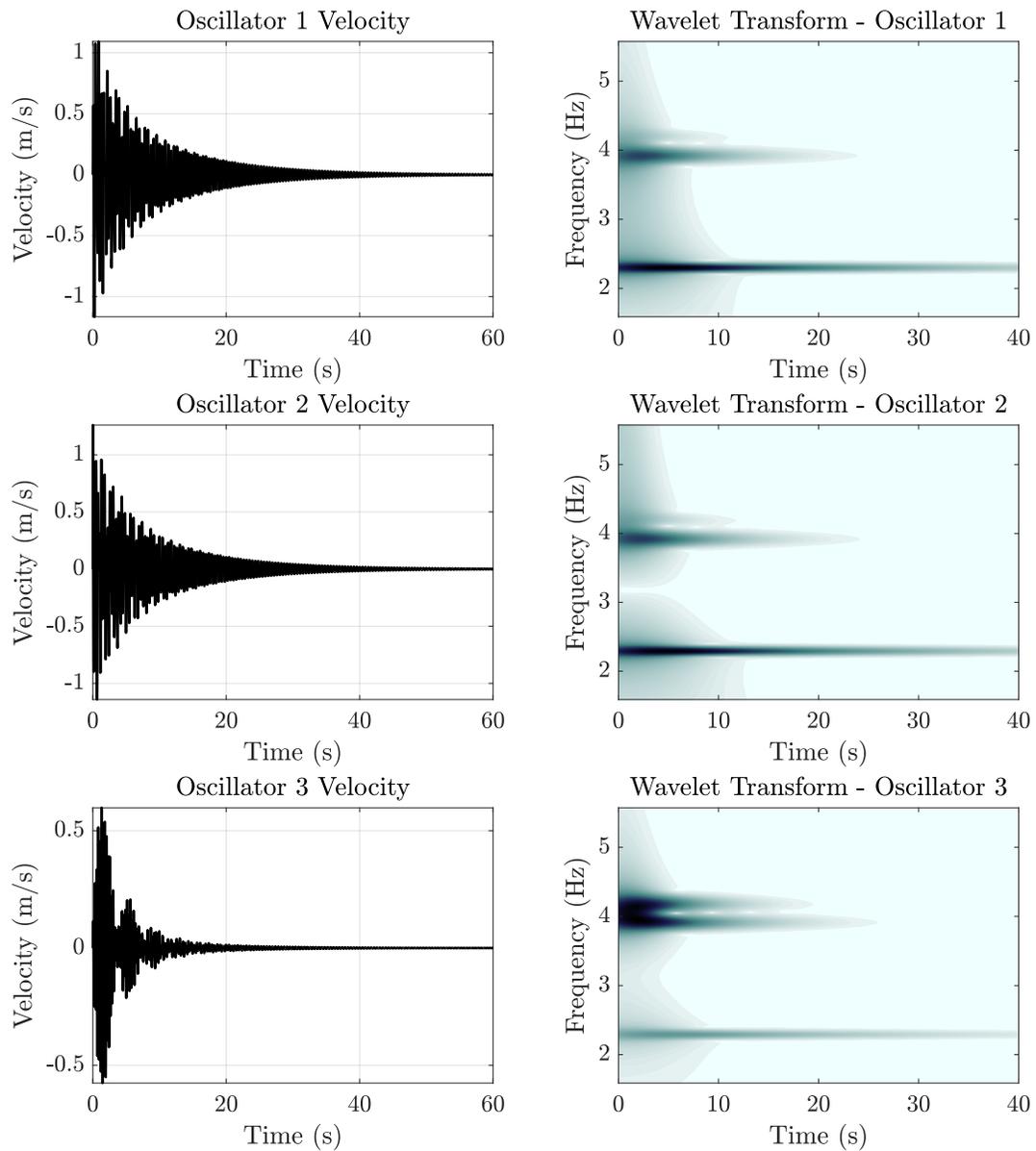} %
    \caption{Drive point at oscillator 2; time-frequency representation of the system response using the CWT.}
    \label{fig:cwt_response_2}
\end{figure}

\begin{figure}[H]
    \centering
    \includegraphics[width=1\textwidth]{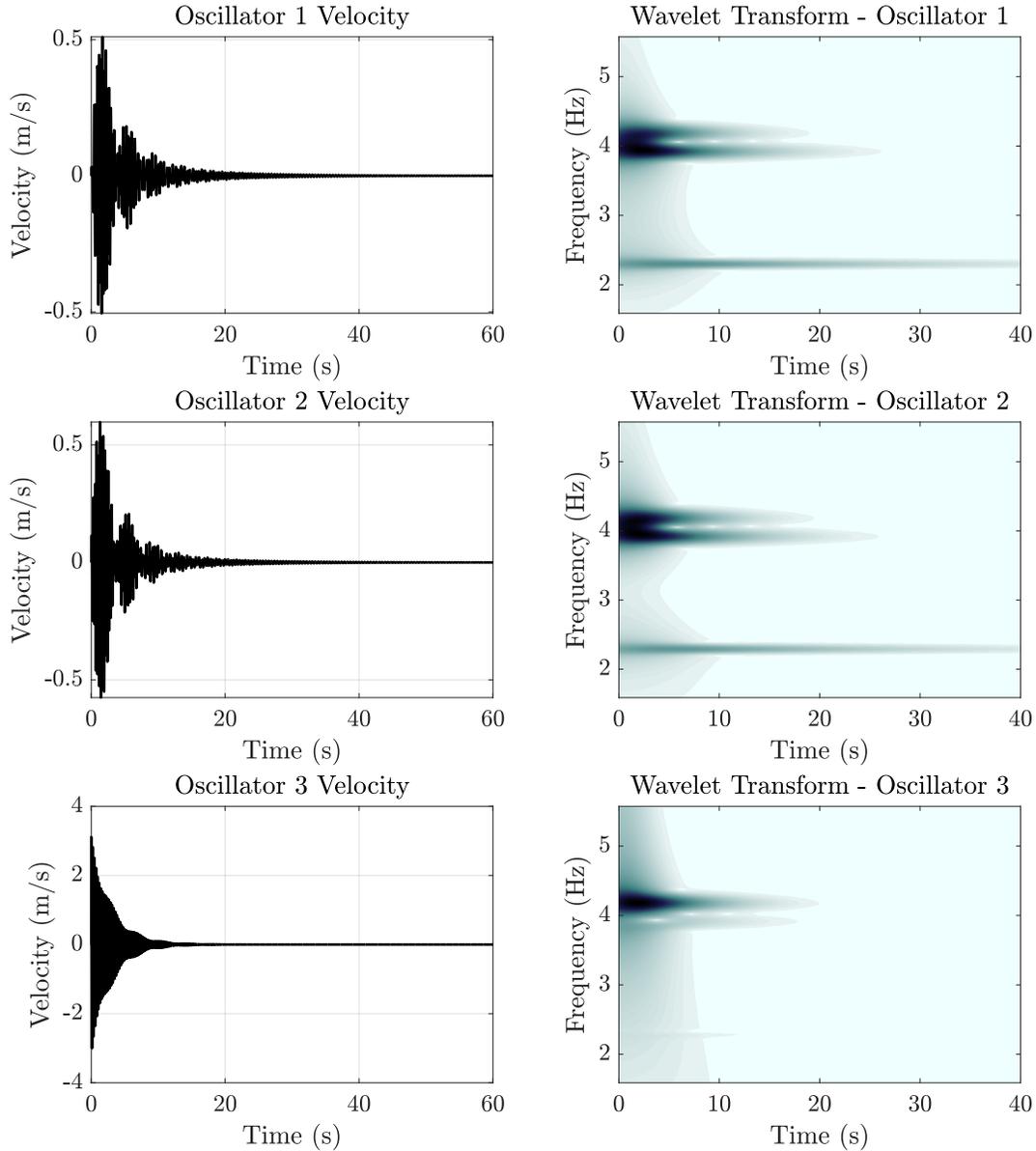} 
    \caption{Drive point at oscillator 3; time-frequency representation of the system response using the CWT.}
    \label{fig:cwt_response_3}
\end{figure}

The matrices \( \mathbf{D} = \mathbf{M}^{-1}\mathbf{C} \) and \( \mathbf{S} = \mathbf{M}^{-1}\mathbf{K} \) represent the mass-normalized damping and stiffness matrices, respectively. For this case study, we select the physical parameters as $m = 1\, \text{kg}, \quad k = 100\, \text{N/m}, \quad c = 0.1\, \text{Ns/m}, \quad \epsilon = 0.1, \quad k_3 = 100(3 - \sqrt{3})\, \text{N/m}.$ The damping matrix incorporates additional localized and asymmetric contributions, defined by
$c_1 = 0.01 \text{Ns/m}, \quad c_2 = 0.02 \text{Ns/m}, \quad c_3 = 0.01 \text{Ns/m}, \quad c_{2A} = 0.01 \text{Ns/m}, \quad c_{3A} = 0.08$ \text{Ns/m}. As a result, the third oscillator (associated with displacement $x_3$) is tuned such that its natural frequency matches that of the lower mode of the two-DOF subsystem which is obtained in the zero coupling limit $\epsilon \to 0$. This ensures that for a small but nonzero $\epsilon$, the full three-oscillator system exhibits two closely spaced modes resulting from near-resonant modal coupling. Furthermore, this modeling choice ensures that the damping matrix $\mathbf{C}$ is such that it breaks the classical damping distribution assumption commonly used in classical modal analysis \cite{ewins, caughey1965classical}. The resulting system possesses complex modes, with non-trivial phase differences between modal coordinates.
\vspace{0.5em}
 The system is subjected to transient external excitation applied to one of the masses modeled as,
\begin{equation}
\mathbf{M}^{-1}f(t) =
\begin{cases}
f_0 \sin\left(\frac{\pi t}{t_d}\right), & 0 \leq t \leq t_d, \\
0, & t > t_d,
\end{cases}
\end{equation}
with $f_0 = 10\, \text{N/kg}$, $t_d = 0.001$. This type of smooth, finite-duration excitation is commonly used to excite low-pass frequency content and is particularly useful in identifying multiple modes from the system's transient response. The resulting responses of system ~\eqref{eq:gov_eq} are analyzed using the CWT, as shown in Figs.~\ref{fig:cwt_response_1},~\ref{fig:cwt_response_2},~\ref{fig:cwt_response_3} for three different drive point cases; the results highlight a clear beat phenomenon arising between the two closely spaced modes, which appears as modulated interference in the time-frequency representation and creates additional complexity in the harmonic decomposition and modal separation processes.

\subsection{System Identification}

\begin{figure}[H]
    \centering
    \includegraphics[width=0.5\textwidth]
    {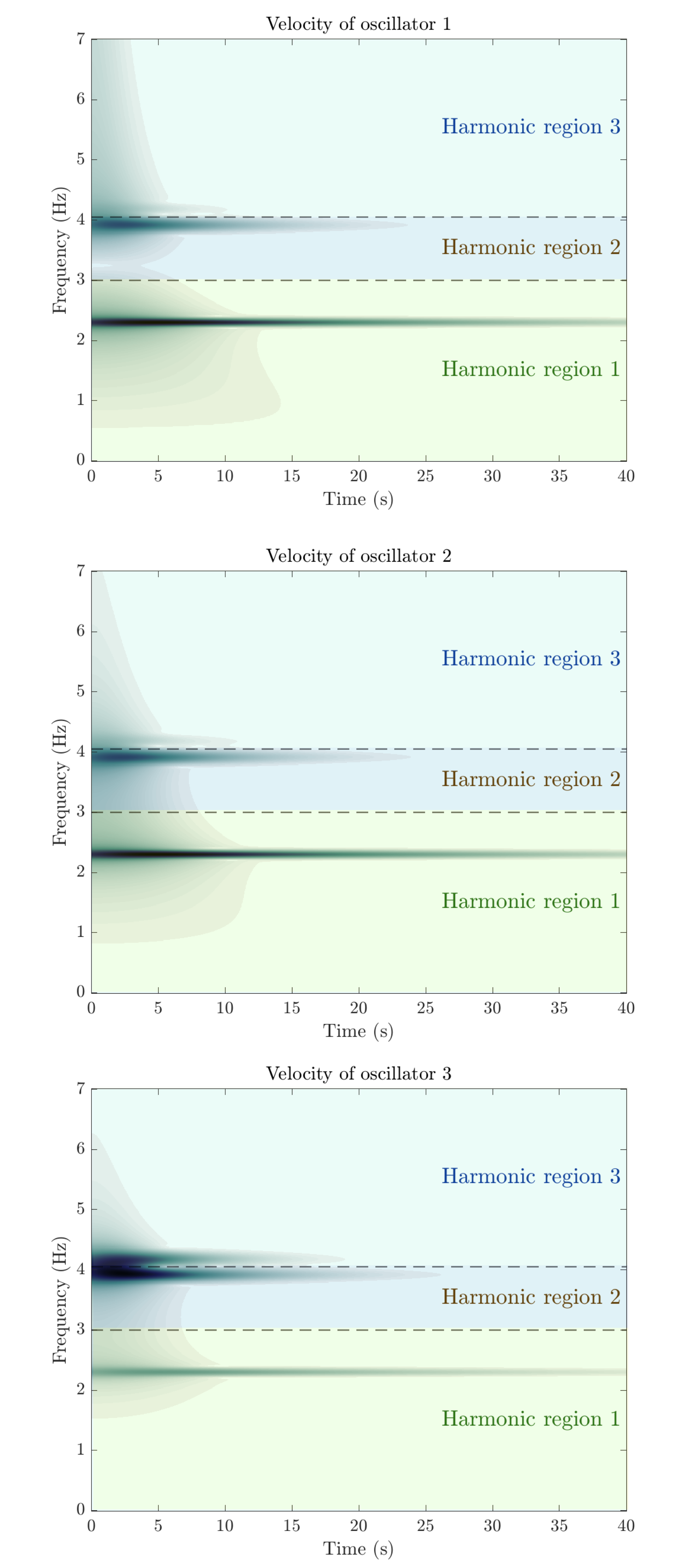}
    \caption{Selected harmonic regions for the ICWT computations - driving point at oscillator 1.}
    \label{fig:harmonic_regions}
\end{figure}

Following the CWT analysis, the outlined wavelet-based identification protocol is applied to extract the dominant modal contributions from the simulated system response. 
We explicitly illustrate the process of deriving the harmonic time series through the scenario of impulse excitation (drive point) at oscillator 1. Starting from the raw measured velocity response  (Fig.~\ref{fig:cwt_response_1}), and its CWT spectrum we select the harmonic regions that are then identified based on the topology of this spectrum (see Fig.~\ref{fig:harmonic_regions}); note that by selecting these harmonic regions coherent wavelet results corresponding to modal harmonics are visually separated. These regions are chosen to encompass the localized energy concentrations associated with the fundamental modal frequencies, while minimizing as much as possible interference from modal overlap.
% {figures2/vertial_wav_num.png}

\begin{figure}[H]
    \centering
    \includegraphics[width=1\textwidth]{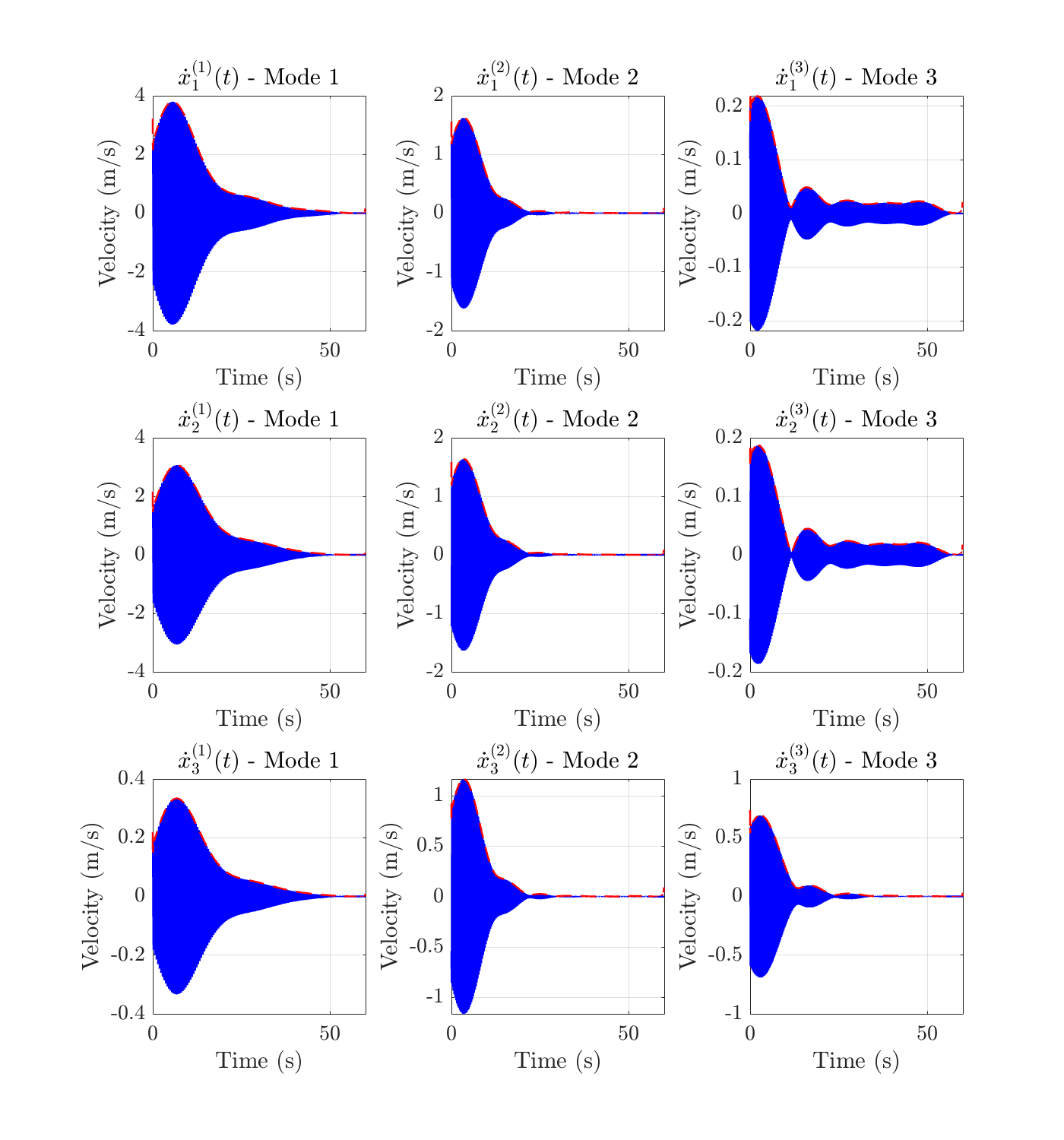}
    \caption{Harmonic decomposition of the velocities for drive point at oscillator 1, with analytic envelopes (red dashed lines) obtained by Hilbert transform.}
    \label{fig:icwt_components}
\end{figure}

After defining the harmonic regions, the ICWT expression Eq.~\eqref{eq:icwt_direct} is applied independently to each region. In this way, we obtain the decomposed time-domain signals $\dot{x}_i^{(j)}(t)$, corresponding to the $j$-th harmonic component of each measured velocity response $\dot{x}_i(t)$. These harmonic components represent the modal contributions embedded within the original raw signals and are illustrated in Fig.~\ref{fig:icwt_components}. Hence, we compute the modal contributions in the time domain for each of the velocity responses of system ~\ref{eq:gov_eq} when an impulse is applied to oscillator 1. Similar transient decompositions of the velocities of the system for driving points at oscillators 2 (Fig. ~\ref{fig:cwt_response_2}) and 3 (Fig. ~\ref{fig:cwt_response_3}) are obtained (not shown here).

Returning to the results of Fig.~\ref{fig:icwt_components}, by applying the Hilbert transform to each transient modal time series $\dot{x}_i^{(j)}(t)$ we extract their analytic representations. This enables direct computation of both their moduli and instantaneous phases. The envelopes provide reliable estimates of the corresponding modal amplitudes, while the instantaneous phases capture the relative phase differences between the degrees of freedom. These two quantities are then combined to construct complex-valued mode shapes, as explained in detail in Subsection~\ref{sec:wavelet_protocol}.
This step is essential when analyzing systems with non-classical damping and closely spaced modes, since real-valued mode shapes alone fail to capture the underlying dynamics. The ability to introduce complex-valued mode shapes from raw signals forms a central contribution of the proposed methodology.

\begin{figure}[htbp]
    \centering
    \includegraphics[width=1\textwidth]{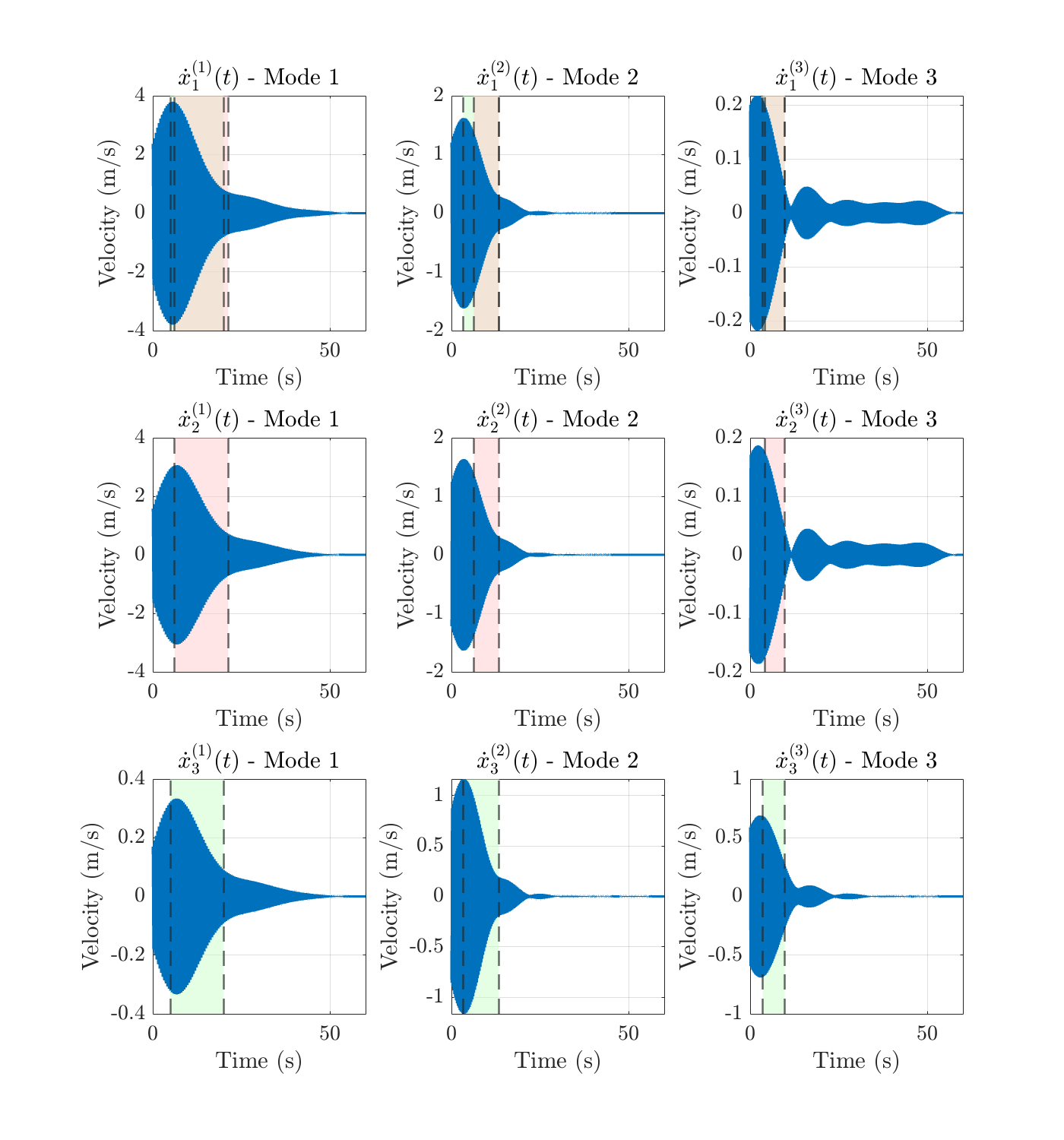}
    \caption{Harmonic decomposition of the impulse response with excitation applied at oscillator~1. Green and red shaded regions indicate overlapping time windows selected for reliable phase difference estimation between oscillators 3 and 1, and between oscillators 2 and 1, respectively.}
    \label{fig:hilbert}
\end{figure}

Indeed, the ICWT is applied in combination with the Hilbert transform to extract the moduli and phase differences that are necessary for estimating the identified complex-valued modal matrix. Figure~\ref{fig:hilbert} illustrates the harmonic decomposition for the numerical case study, where an impulse excitation was applied to oscillator~1. Each subplot displays the reconstructed mode obtained via ICWT. The accurate recovery of the amplitude envelopes demonstrates the efficacy of the Hilbert transform for instantaneous amplitude estimation.

To reliably estimate the modal phase differences, we select appropriate overlapping time windows. These windows are highlighted in Figure~\ref{fig:hilbert} using different colors: the green-shaded region indicates the time segment chosen for comparing the decomposed modal velocities of oscillator~3 against those of the reference oscillator~1, while the red-shaded region marks the interval selected for the respective comparison for the modal velocities of oscillator 2. As described in detail in Section~\ref{sec:wavelet_protocol}, each overlapping window is selected based on two key criteria: (i) The envelope of the decomposed harmonic component within the window must be sufficiently smooth and maintain adequate amplitude; and (ii) the chosen window must span a common time interval shared by the reference velocity time series and the oscillator velocity time series under analysis. 

Considering that for a given applied impulse some of the oscillators of the system may be strongly excited, we conduct the phase estimation procedure across multiple measurement sets. Subsequently, the optimization procedure described in Section~\ref{sec:drive_point_cases_velocity} robustly averages these individual phase estimates, thereby yielding accurate and reliable complex-valued mode shapes. The identified modal parameters obtained via this data-driven protocol are summarized and compared against the exact values computed from the known system matrices in Table~\ref{tab:modal_comparison} and ~\ref{tab:modal_comparison2}.
The resulting satisfactory agreement validates the effectiveness of the proposed method in resolving closely spaced complex modes without relying on prior knowledge of the governing equations; rather the entire system identification is purely data driven. 

\begin{figure}[H]
    \centering
    \includegraphics[width=1\textwidth]{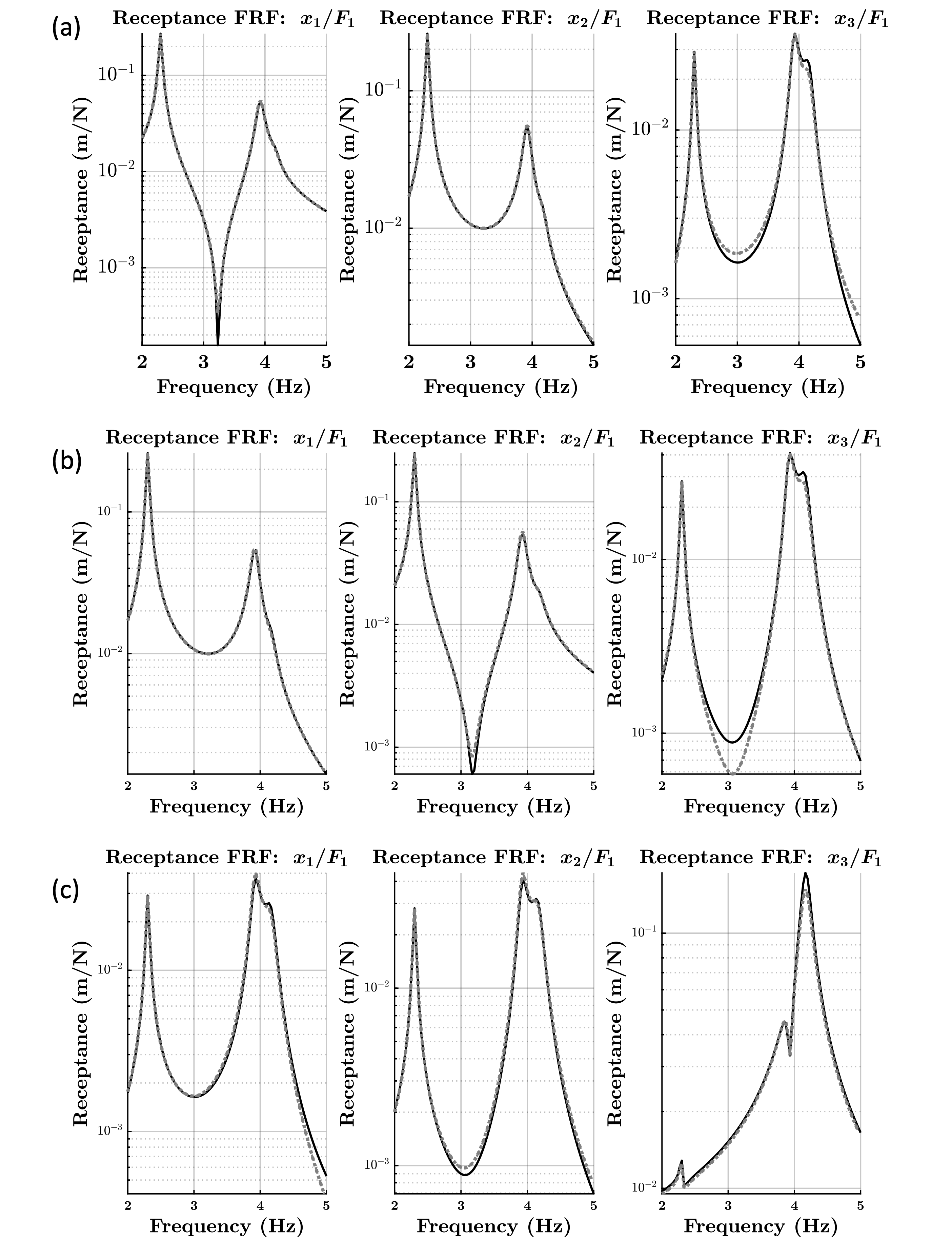}
    \caption{Comparison between exact numerical (black lines) and reconstructed (gray lines) FRFs for drive-point excitations applied at (a) oscillator 1, (b) oscillator 2, and (c) oscillator 3.}
    \label{fig:frf_reconst_numerical}
\end{figure}

Furthermore, Fig.~\ref{fig:frf_reconst_numerical} illustrates the reconstructed FRFs for varying drive point excitations at oscillators 1, 2, and 3, respectively. This figure highlights the satisfactory agreement between the numerically derived ground-truth FRFs and those reconstructed using the proposed wavelet-based identification procedure. Notably, the reconstruction accurately captures the intricate valleys between the two closely spaced modes, a result made possible only by properly accounting for mode shape complexification introduced through instantaneous phase estimation.

\subsection{Wavelet-Based ROM Reconstruction}

\begin{figure}[H]
    \centering
    \includegraphics[width=0.6\textwidth]{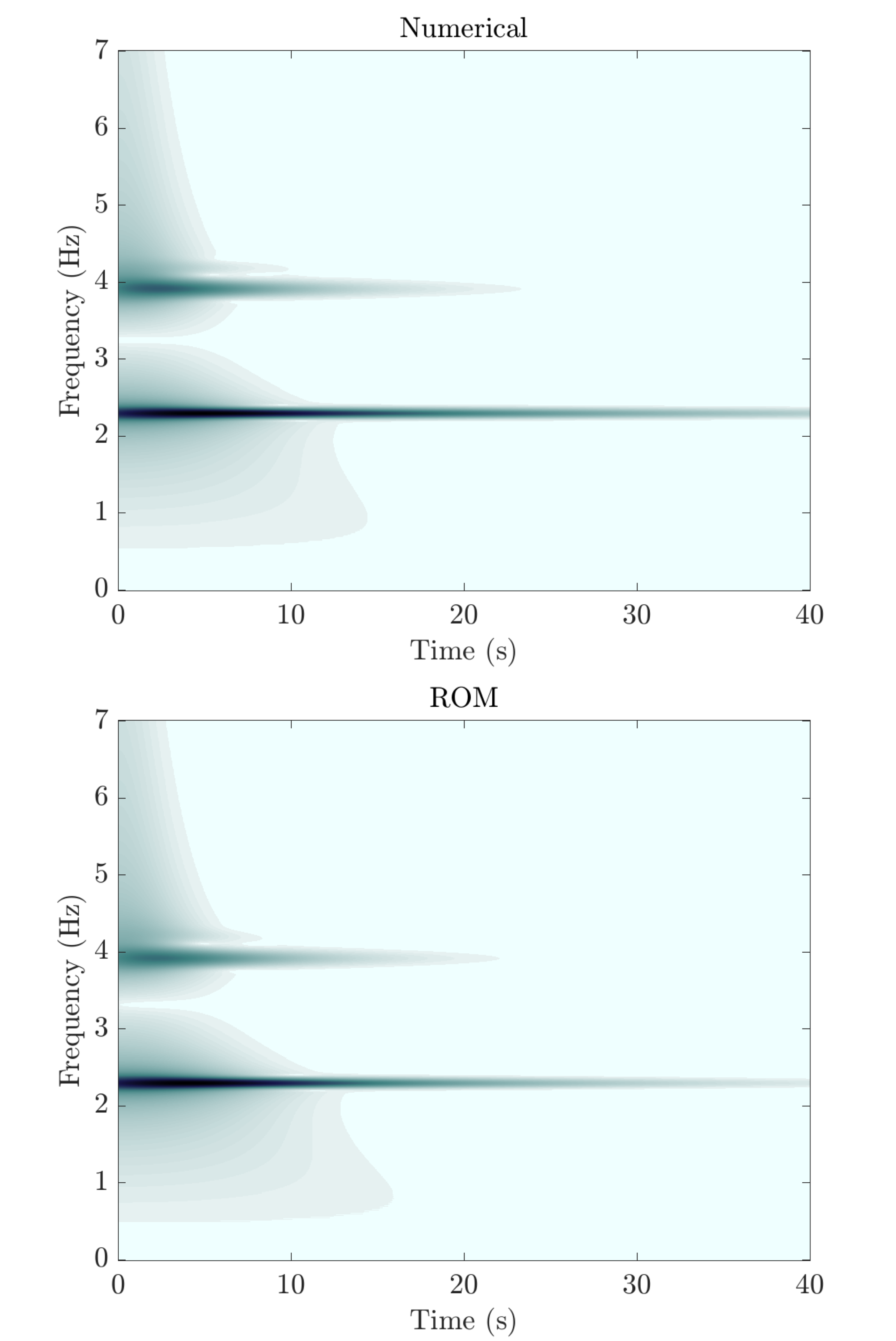}
    \caption{Comparison between measured (top) and ROM reconstructed (bottom) continuous wavelet spectra of the velocity of oscillator 1 for impulse excitation applied at oscillator 1.}
    \label{fig:rom_validation}
\end{figure}

As a final step, we reconstruct the CWT of the responses using a ROM developed on wavelet transforms as discussed in Section~\ref{sec:rom_reconstruction}. By applying to the ROM an excitation identical to the one applied to oscillator 1 in the previous exact simulation, we replicate accurately the system's dynamics in time-frequency. Presented in Figure~\ref{fig:rom_validation} is a qualitative comparison between the numerically-obtained continuous wavelet spectrum (top) and the ROM-based reconstruction (bottom). The agreement further confirms the accuracy of the identified modal parameters and the efficacy of the proposed wavelet-based system identification framework.

\begin{table}[H]
    \centering
    \caption{Comparison of analytically-derived vs. wavelet-identified modal parameters I.}
    \label{tab:modal_comparison}
    \begin{tabular}{cccccc}
        \hline
        Mode & Quantity & Analytical & Identified & Relative Error & Units \\
        \hline
	\multirow{2}{*}{1}
        & Frequency (\( f_1 \)) & 2.30 & 2.30 & 0\% & Hz \\
        & Damping (\( \zeta_1 \)) & 0.91\% & 0.91\% & 0.25\% & -- \\
        \hline
        \multirow{2}{*}{2}
        & Frequency (\( f_2 \)) & 3.92 & 3.92 & 0\% & Hz \\
        & Damping (\( \zeta_2 \)) & 1.54\% & 1.48\% & 3.65\% & -- \\
        \hline
        \multirow{2}{*}{3}
        & Frequency (\( f_3 \)) & 4.17 & 4.17 & 0\% & Hz \\
        & Damping (\( \zeta_3 \)) & 2.00\% & 2.25\% & 12.28\% & -- \\
        \hline
    \end{tabular}
\end{table}

\begin{table}[H]
\centering
\caption{Comparison of exact and estimated modal parameters II.}
\label{tab:modal_comparison2}
\begin{tabular}{c|ccc|ccc}
\toprule
Oscillator & \multicolumn{3}{c|}{\textbf{Exact Relative Phase (deg)}} & \multicolumn{3}{c}{\textbf{Estimated Relative Phase (deg)}} \\
\hline
1 & 0.000 & 0.000 & 0.000 & 0.000 & 0.000 & 0.000 \\
2 & 1.002 & 178.743 & -164.275 & 1.108 & 178.668 & -165.709 \\
3 & 18.977 & 138.048 & 51.900 & 15.353 & 140.087 & 45.720 \\
\hline
 & \multicolumn{3}{c|}{\textbf{Exact Moduli}} & \multicolumn{3}{c}{\textbf{Estimated Moduli}} \\
\hline
1 & 0.720 & 0.621 & 0.143 & 0.724 & 0.627 & 0.145 \\
2 & 0.689 & 0.625 & 0.163 & 0.685 & 0.644 & 0.176 \\
3 & 0.077 & 0.473 & 0.976 & 0.077 & 0.429 & 0.971 \\
\bottomrule
\end{tabular}
\end{table}

\section{Experimental Study: Airplane Model}

To further validate the robustness and accuracy of the proposed methodology, we consider an experimental case study. Specifically, we experimentally demonstrate the application of the system identification methodology to a model airplane with closely spaced modes.

\subsection{Experimental Setup}
\label{subsec:experimental_setup}

\begin{figure}[H]
    \centering
    \includegraphics[width=1\textwidth]{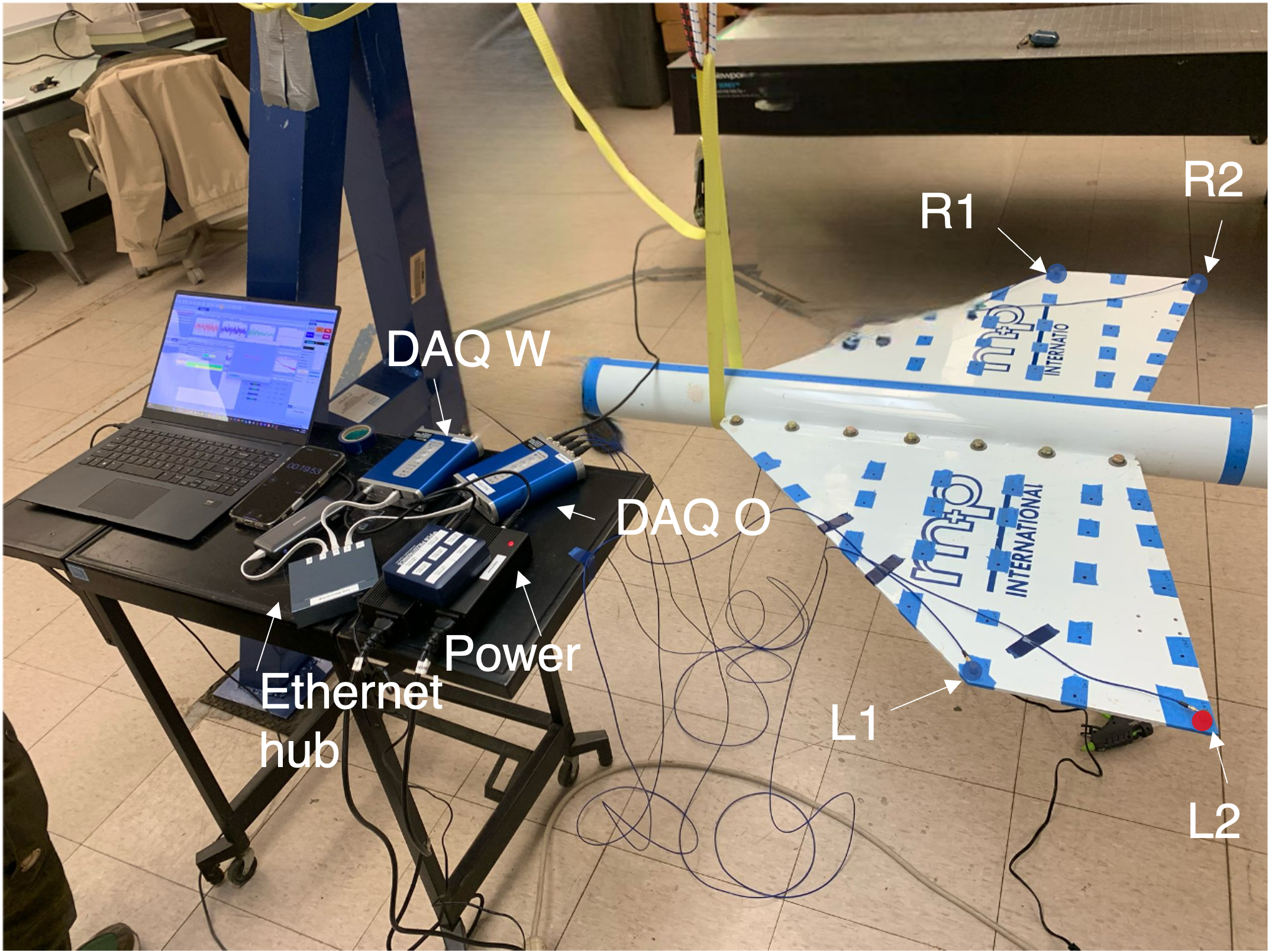} 
    \caption{Experimental setup for wavelet-based modal testing using an airplane model suspended by bungees. Accelerometers are positioned at the wingtips (L1, L2, R1, R2), and vertical excitation is provided via impact hammer at each measurement location.}
    \label{fig:experimental_setup}
\end{figure}

The methodology was experimentally validated using a steel airplane model exhibiting closely spaced modal frequencies. The experimental fixture is depicted in Figure~\ref{fig:experimental_setup}. An airplane model was suspended using bungees to approximate free-free boundary conditions, minimizing external constraints and mimicking idealized in-flight boundary conditions. More details about the dimensions and composition of the experimental model airplane can be found in \cite{Chang2025}. The test procedure involved impact hammer testing to excite the structure and collect response data. Acceleration time series were measured at four wingtip locations labeled L1, L2, R1, and R2 in Fig.~\ref{fig:experimental_setup} using PCB Piezotronics uniaxial accelerometers. The impulsive load at each of these four locations was applied in the vertical direction. Data acquisition was performed using two synchronized Crystal Instruments Spider-20HE data acquisition systems and Engineering Data Management software. At each wingtip location, a PCB Piezotronics impact hammer was used to apply and measure the force input, and acceleration time series at each location was measured. This setup allowed comprehensive capturing of the system’s dynamic characteristics necessary for the identification of closely spaced modes.

The recorded experimental data were processed following the protocol described in previous sections, leading to accurate extraction of modal parameters and subsequent validation through reconstruction of the dynamics.

\subsection{Results of System Identification}

\begin{figure}[H]
    \centering
    \includegraphics[width=0.8\textwidth]{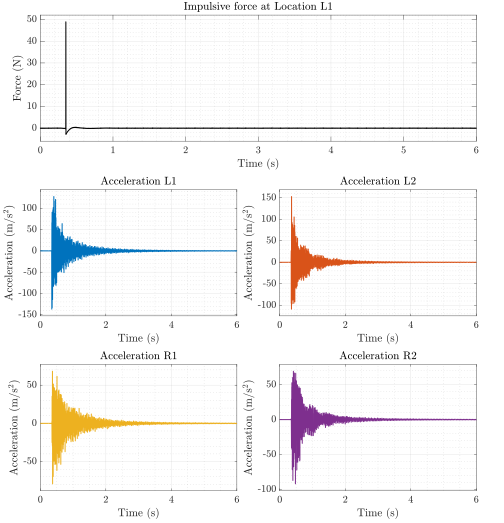} 
    \caption{Measured input force and acceleration responses at selected locations of the airplane model. The top plot shows the hammer input force (N) at Location L1, while the bottom four plots show acceleration responses at the sensing locations L1, L2, R1, and R2 (m/s²).}
    \label{fig:experimental_l1}
\end{figure}

Representative experimental time series at the four measurement locations for excitation applied at location L1 are presented in Figure~\ref{fig:experimental_l1}. Two independent data sets per drive point were collected to ensure repeatability and reliability in the measurements. A typical hammer input force is shown at the top, while the corresponding structural responses at the four sensor locations are displayed below. These acceleration signals form the basis for the subsequent wavelet-based system identification procedure.

Given the broadband nature of the excitation, we focus our analysis on the frequency regime between 100–200 Hz. This range is of particular interest because it contains two closely spaced modes, together with two more isolated modes, which present a natural challenge for reliable identification. To disentangle these contributions, we apply our data-driven wavelet framework.

\begin{figure}[H]
    \centering
    \includegraphics[width=0.65\textwidth]{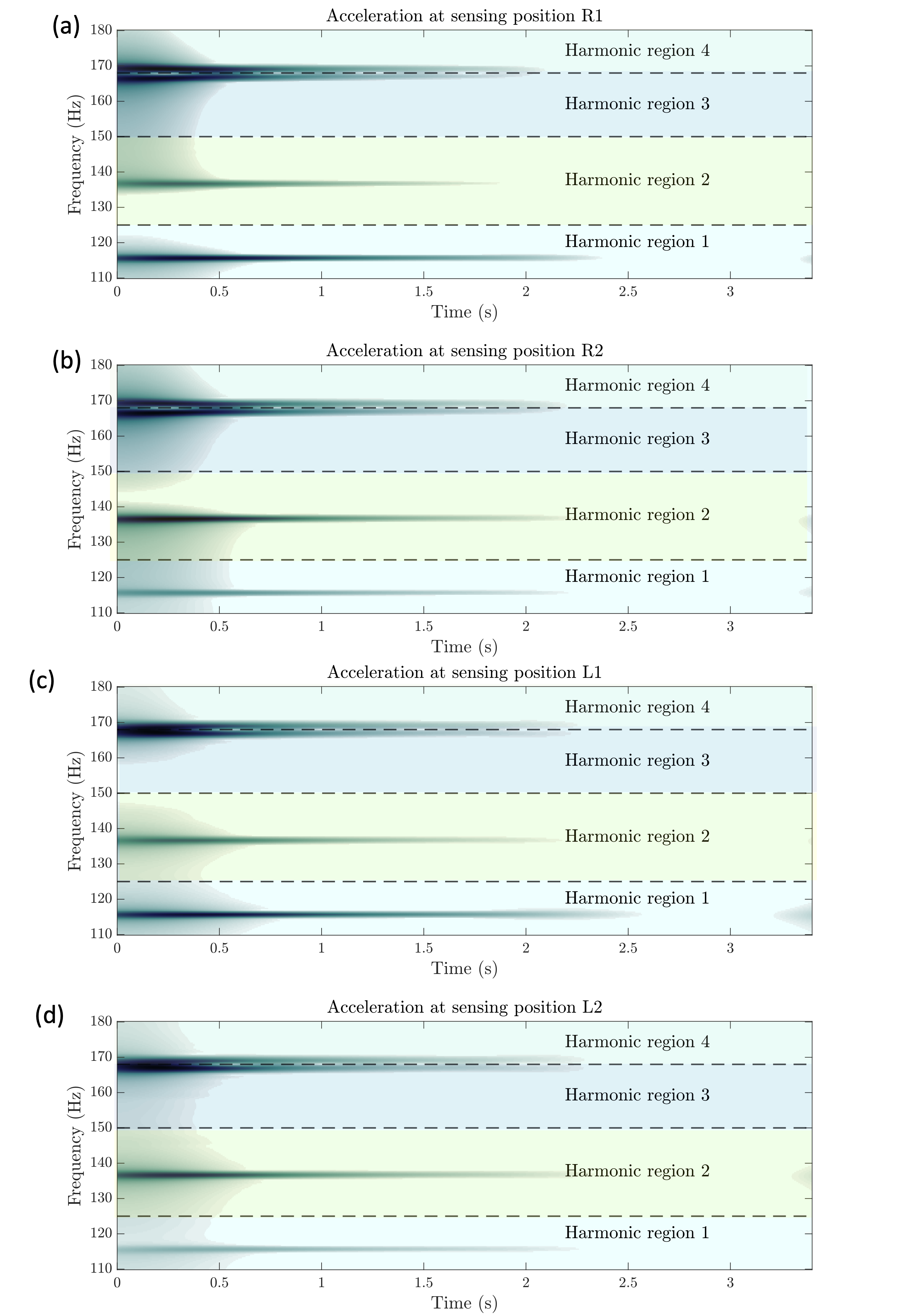} 
    \caption{CWT spectra of the measured accelerations showed in Fig.~\ref{fig:experimental_l1}, at locations (a) R1, (b) R2, (c) L1 and (d) L2. The separation of each wavelet spectrum into four harmonic regions is depicted. Each selected band corresponds to a separate harmonic region (labeled 1–4), within which the corresponding modal contributions are extracted via the inverse CWT.}
    \label{fig:wav_airplane}
\end{figure}

Figure~\ref{fig:wav_airplane} illustrates the CWT spectra of the accelleration time series at the four sensor locations (see Fig.~\ref{fig:experimental_l1}). The energy distribution in the time–frequency plane clearly reveals the modal content of the structural responses. Horizontal dashed lines are used to separate the frequency axis into four distinct harmonic regions. Each shaded region isolates the contribution of a single mode, thus providing a systematic means to capture the closely spaced modal dynamics. Within each region, the ICWT is applied to reconstruct the time-domain modal responses.

\begin{figure}[H]
    \centering
    \includegraphics[width=1\textwidth]{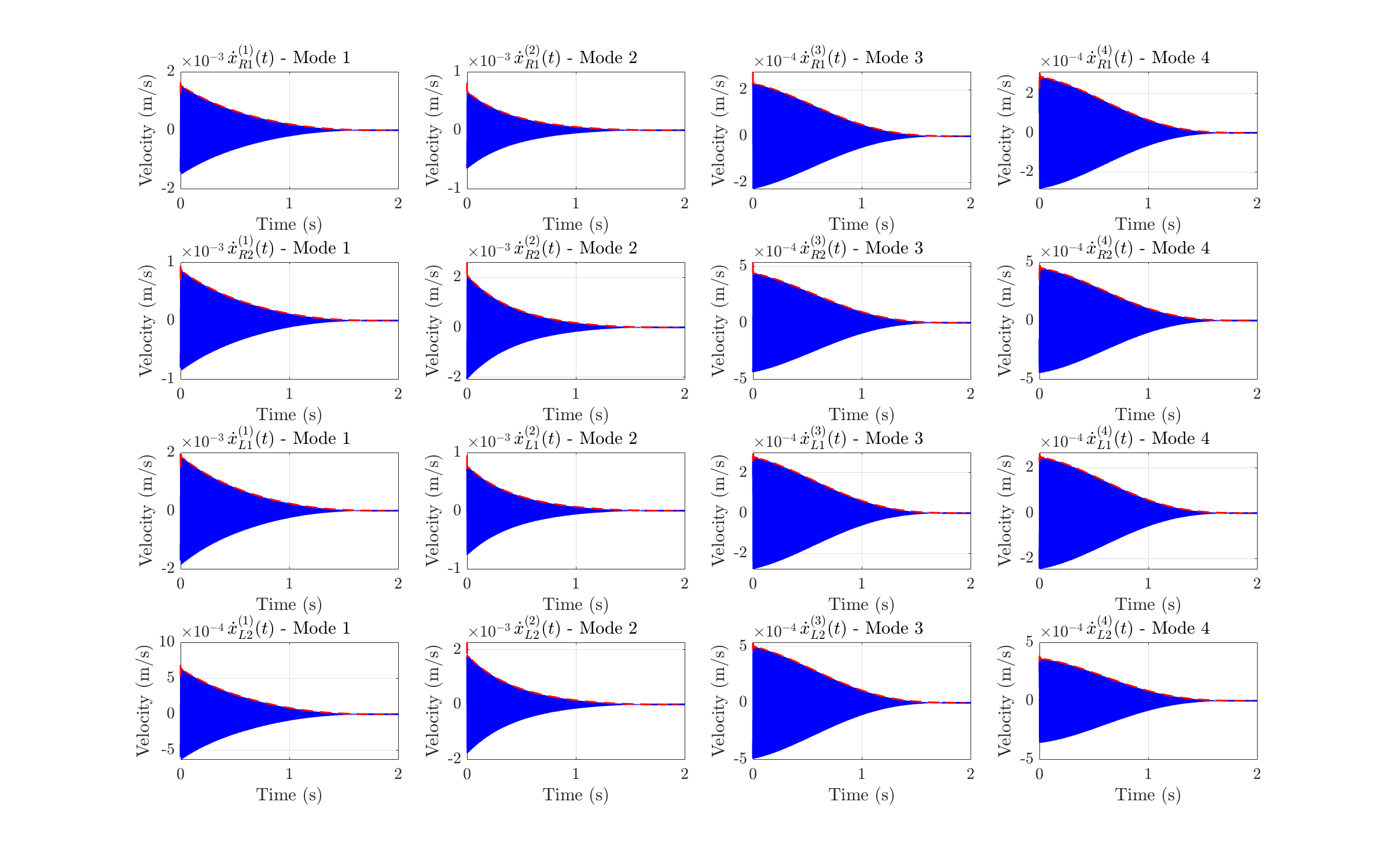} 
    \caption{Postprocessing of measured data for impulsive excitation applied at Location L1: Decomposed harmonic components of the velocity time series (blue) with superimposed analytical envelopes (red dashed) for the airplane model at four measurement locations (R1, R2, L1, L2) across four extracted modes. The envelopes are truncated at 1.95 s to emphasize the initial modal decay behavior.}
    \label{fig:envelopes_airplane}
\end{figure}

\begin{figure}[H]
    \centering
    \includegraphics[width=1\textwidth]{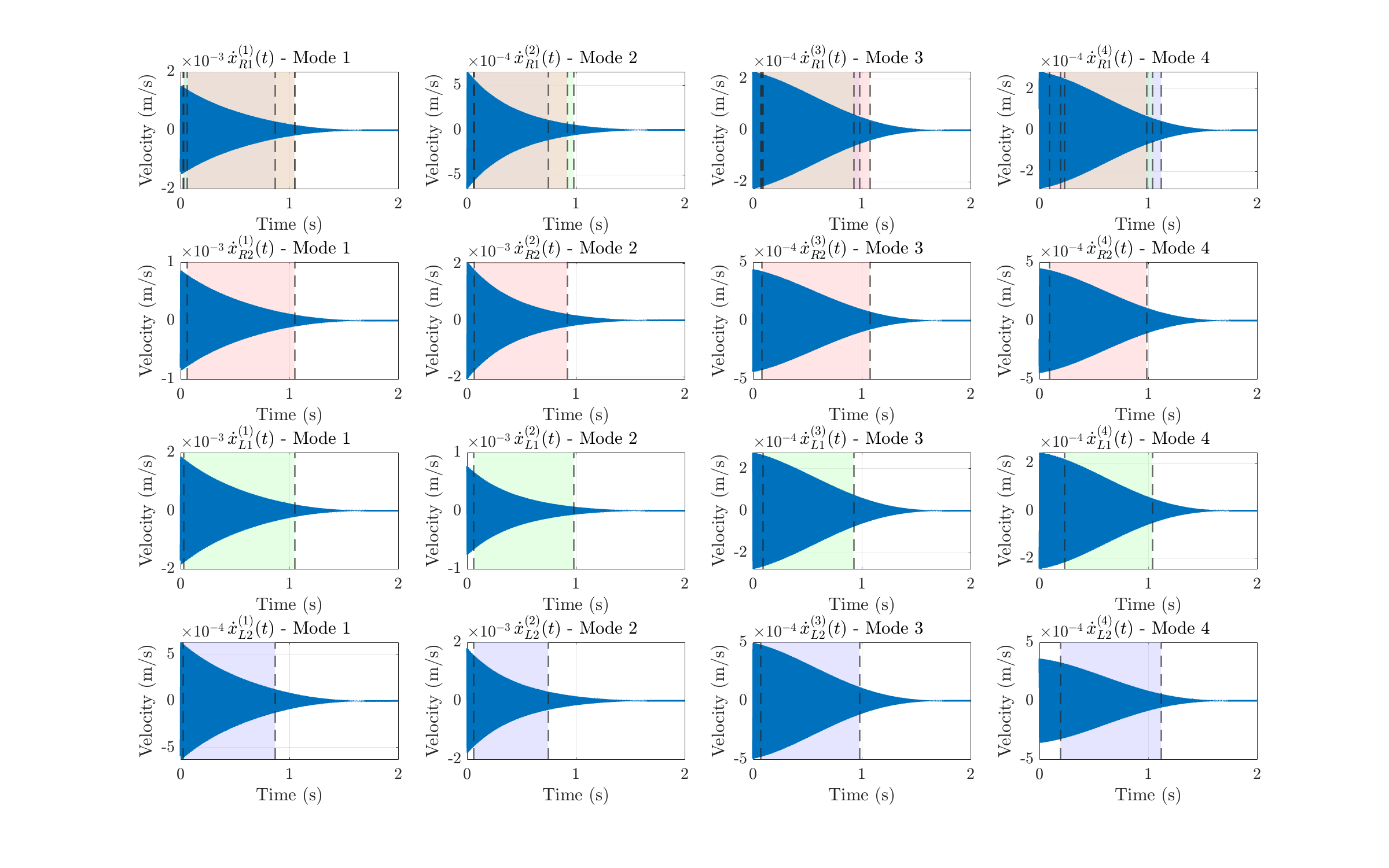} 
    \caption{Postprocessing of the decomposed velocity components depicted in Fig. 14 to derive the modal complexities: Velocity time series depicted in Fig. ~\ref{fig:envelopes_airplane}, shaded windows indicate the intervals selected for phase-difference analysis. These windows were used to determine relative phase shifts and construct the complex mode shapes.}
    \label{fig:windows_airplane}
\end{figure}

As in the previous section, the reconstructed signals are further analyzed by extracting their analytical envelopes via the Hilbert transform. As shown in Figure~\ref{fig:envelopes_airplane}, the velocity time series at each location (obtained by numerically integrating the measured accelerations shown in Fig.~\ref{fig:experimental_l1}), are superimposed to their corresponding analytical envelopes. To capture the complex modal behavior, we perform an overlapping time-window selection of the reconstructed responses. This step, illustrated in Figure~\ref{fig:windows_airplane}, allows us to quantify rhe phase differences between the measured velocities at different locations (R1, R2, L1, L2) for each mode. By combining the amplitude envelopes obtained from the Hilbert envelopes with the phase-difference information extracted from the overlapping windows, we construct the complex-valued mode shapes of the structure.

Table~\ref{tab:exp_modal_matrix} summarizes the identified modal parameters, including natural frequencies and damping ratios derived through our wavelet-based methodology. Furthermore, Table~\ref{tab:exp_modal_matrix} presents the corresponding mass-orthonormalized modal matrix obtained by applying the system identification protocol described in previous sections.

\begin{figure}[H]
    \centering
    \includegraphics[width=1\textwidth]{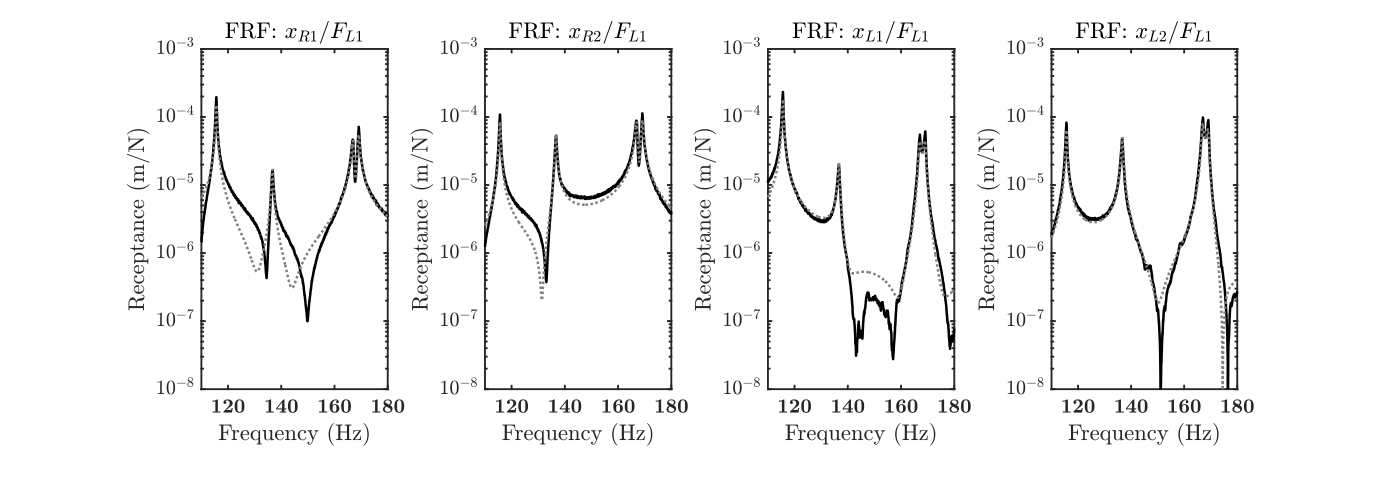} 
    \caption{Comparison of experimental (black lines) and reconstructed (gray lines) FRFs for all four measurement points. Excitation is applied at location L1.}
    \label{fig:frf_reconstruction_L1}
\end{figure}

As an additional validation, we reconstruct the FRFs using the identified modal parameters. The resulting FRF reconstructions are compared with the experimentally measured FRFs in Figure~\ref{fig:frf_reconstruction_L1}. The agreement between the reconstructed and experimental FRFs is satisfactory, particularly around the closely spaced modes. A minor mismatch is observed at lower amplitude regions, which is attributed to measurement noise and inherent nonlinearities but does not detract from the validity or effectiveness of the methodology.

\begin{table}[H]
\centering
\caption{Experimentally Identified Modal Parameters: Frequencies, Damping Ratios, Relative Angles, and Moduli}
\renewcommand{\arraystretch}{1.3}
\begin{tabular}{c|cccc}
\hline
\textbf{Mode} & \textbf{1} & \textbf{2} & \textbf{3} & \textbf{4} \\
\hline
\textbf{Frequency (Hz)} & 115.62 & 136.68 & 167.00 & 169.05 \\
\textbf{Damping Ratio \(\zeta\)} & 0.00179 & 0.00205 & 0.00129 & 0.00117 \\
\hline
\multicolumn{5}{c}{\textbf{Relative Phase Angle (degrees)}} \\
\hline
Measurement Point R1 & 0  & 0   & 0   & 0 \\
Measurement point R2 & -0.035 & 1.469   & 179.526 & -179.729 \\
Measurement Point L1 & 0.470  & -178.007 & -3.669  & 179.543 \\
Measurement Point L2 & -0.368 & -178.350 & 177.394 & 0.652 \\
\hline
\multicolumn{5}{c}{\textbf{Mass-Orthonormalized Moduli}} \\
\hline
Measurement Point R1 & 0.581 & 0.218 & 0.316 & 0.400 \\
Measurement Point R2 & 0.327 & 0.688 & 0.595 & 0.649 \\
Measurement Point L1 & 0.702 & 0.255 & 0.332 & 0.344 \\
Measurement Point L2 & 0.248 & 0.626 & 0.608 & 0.505 \\
\hline
\end{tabular}
\label{tab:exp_modal_matrix}
\end{table}

\begin{figure}[H]
    \centering
    \includegraphics[width=1.1\textwidth]{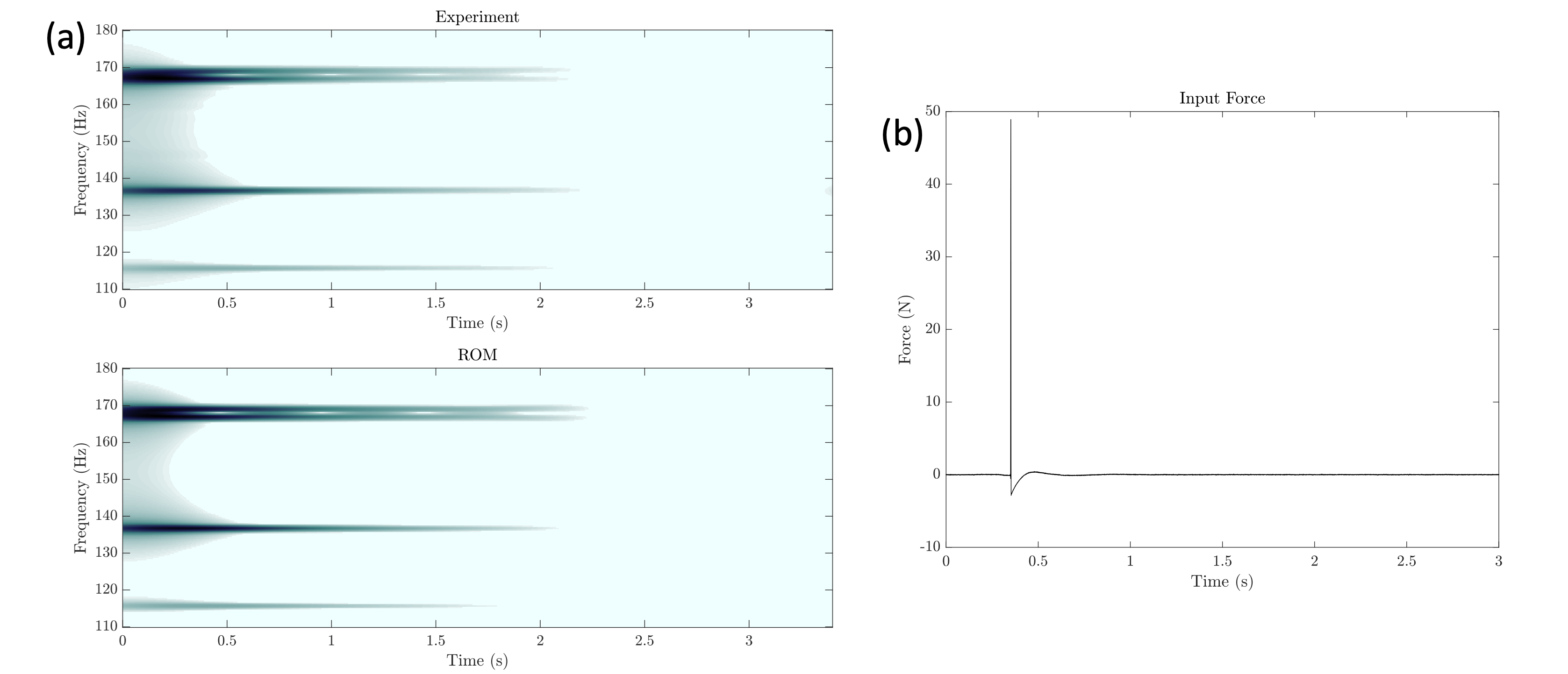}
    \caption{(a) Time–frequency validation of the ROM based on the data-driven wavelet-based identification method. The top panel shows the experimentally obtained CWT spectrum from acceleration response at location L1. The bottom panel shows the CWT of the reconstructed response using the identified ROM, with impact hammer excitation used as input. The excellent agreement in both frequency and temporal resolution between the two confirms the validity and robustness of the identified modal parameters; (b) Applied Impulse at location L2.}
    \label{fig:cwt_rom_validation_L1}
\end{figure}

Finally, we employ a four-DOF ROM to reproduce the continuous wavelet spectrum of the structure under investigation, as measured experimentally in the frequency range of interest. By utilizing the experimentally-recorded impact forcing as input excitation to the ROM, we demonstrate that the ROM captures both temporal evolution and frequency content of the system responses accurately. Figure~\ref{fig:cwt_rom_validation_L1} illustrates the comparison between the experimentally measured and ROM-predicted wavelet spectra. Specifically, we replicate the wavelet spectrum of the time series at location R1 for the case study of excitation being applied to location L1. This strong agreement underscores the robustness and precision of the proposed wavelet-based methodology, emphasizing its capability to yield mass-orthonormalized modal matrices and accurate reduced-order modeling.

\section{Conclusions}

In this work, we have formulated and validated a purely data-driven, wavelet-based methodology for system identification and reduced-order modeling of mechanical structures exhibiting closely spaced modes and classical or non-classical damping. Traditional frequency-domain modal analysis methods typically fall short when resolving complex modal interactions due to limitations inherent in Fourier-based transformations. To address these limitations, we proposed an alternative approach leveraging the powerful joint time–frequency localization capabilities offered by CWT and ICWT. The primary strength of the proposed framework lies in its purely data-driven character, as it requires no prior assumptions about system order or governing equations. By directly extracting modal parameters, including natural frequencies, damping ratios, and complex-valued mass-orthonormalized mode shapes, from measured time series signals, the methodology remains robust even when faced with challenging cases such as closely spaced modes, and non-classical damping.

The wavelet-based identification process implemented here is capable of accurately isolating modal contributions through a strategic harmonic decomposition procedure in the time–frequency domain. Subsequently, the extracted modal signals are analyzed using Hilbert transforms, enabling precise estimation of instantaneous phase and amplitude. The incorporation of overlapping time-window strategies for phase difference estimation further enhances the reliability of complex-valued mode shape reconstruction. The robustness and efficacy of the proposed methodology were demonstrated through both numerical and experimental case studies, validating the results via FRF reconstruction and by replicating the continuous wavelet spectrum using ROMs. In the numerical case study, we investigated a mechanical system characterized by closely spaced modes and non-classical damping, conditions that typically introduce substantial complexity and challenge conventional identification methods. We successfully extracted accurate modal parameters and demonstrated excellent performance in reconstructing the system’s FRFs and reproducing the original wavelet spectrum from the ROM.

For the experimental validation, we employed a steel airplane model exhibiting closely-spaced modes suspended by bungees to approximate free boundary conditions. Data were acquired using accelerometers placed strategically at the wingtips, with excitation provided by an impact hammer. Again, our methodology effectively identified the two naturally occurring closely spaced modes and yielded accurate modal parameters. Validation was performed by reconstructing the measured FRFs and subsequently replicating the experimentally-obtained CWT through a ROM. The strong agreement observed between the measured responses and reconstructed results reinforces the practical applicability, reliability, and precision of our purely data-driven, wavelet-based framework.

A particularly compelling extension involves applying the wavelet-based methodology explicitly to systems characterized by nonlinearities and time-varying properties. Many mechanical and aerospace systems exhibit complex dynamic behaviors such as hardening or softening stiffness characteristics, as well as damping and stiffness properties that evolve over time, resulting in inherently non-stationary signals. Those phenomena are highly non-stationary and introduce significant challenges for traditional Fourier-based approaches, which assume stationarity and cannot capture temporal localization to handle transient or evolving dynamics effectively. The inherent capability of wavelets to simultaneously resolve both temporal and spectral information makes them particularly advantageous in addressing these limitations. Leveraging the enhanced time-frequency resolution offered by wavelets, future work could effectively address these complexities, complementing and potentially surpassing existing methodologies currently employed in engineering practice. Furthermore, incorporating uncertainty quantification methods into the wavelet-based identification process would provide valuable confidence intervals around the estimated modal parameters, significantly enhancing reliability for critical engineering applications. Finally, exploring integration with machine learning techniques—such as neural networks—could further improve predictive modeling capabilities, enabling real-time monitoring, diagnosis, and control of complex engineering systems.

\section*{ACKNOWLEDGMENTS}

Benjamin J. Chang was supported in part by the Naval Air Warfare Center airplane Division (NAWCAD), Naval
Innovative Science and Engineering (NISE) Program. This support is greatly acknowledged. Any opinions, findings, and conclusions or recommendations
expressed in this work are those of the authors and do not necessarily reflect the views of NAWCAD.

\bibliographystyle{elsarticle-num}
\bibliography{references}
% \bibliographystyle{cas-model2-names}
% \biboptions{numbers,sort&compress}
% \bibliography{references}

\end{document}